\documentclass{JHEP3}
\usepackage[dvips]{graphicx}

\newcommand{\ba}{\begin{array}{c}}
\newcommand{\ea}{\end{array}}

\newcommand{\be}{\begin{equation}}
\newcommand{\ee}{\end{equation}}

\newcommand{\mB}{\mathcal{B}}

\newcommand{\mN}{\mathcal{N}}

\title{Chiral dynamics predictions for \mbox{\boldmath $\eta^\prime\to\eta\pi\pi$}}
\author{
Rafel Escribano$^a$, Pere Masjuan$^b$, Juan Jos\'{e} Sanz-Cillero$^c$\\
$^a$Departament de F\'{\i}sica, Universitat Aut\`{o}noma de Barcelona,
E-08193 Bellaterra (Barcelona), Spain\\
$^b$CAFPE and Departamento de F\'{\i}sica Te\'{o}rica y del Cosmos, Universidad de Granada, Campus de Fuente Nueva, E-18002 Granada, Spain\\
$^c$Istituto Nazionale di Fisica Nucleare INFN, Sezione di Bari,
Via Orabona 4, I-70126 Bari, Italy\\
E-mail:
\email{escribano@ifae.es},
\email{masjuan@ugr.es},
\email{juan.sanzcillero@ba.infn.it}
}

\abstract{
The hadronic decays $\eta^\prime\to\eta\pi\pi$ are studied in the frameworks of
large-$N_C$ Chiral Perturbation Theory, at lowest and next-to-leading orders,
and Resonance Chiral Theory in the leading $1/N_C$ approximation.
Higher order effects such as $\pi\pi$ final state interactions are taken into account through
a detailed unitarization procedure.
The inclusion of finite-width effects in the case of RChT is also discussed.
The Dalitz plot distribution and the differential branching ratio are computed in both approaches.
The predicted Dalitz plot parameters obtained from the different treatments are compared with the most recent measured values.
We find that the $\eta^\prime\to\eta\pi\pi$ branching ratios are easily understood,
while the Dalitz plot parameters require the inclusion of $\pi\pi$ loops
in order to achieve a reasonable agreement.
Our final predictions agree with the experimental measurements.
We hope our results to be of relevance for present and future experimental analyses of these decays.
}
\preprint{
UAB--FT--685\\
UGFT-278/10\\
CAFPE-148/10\\
BARI-TH/637-10}

\begin{document}

\section{Introduction}
\label{intro}
Chiral Perturbation Theory (ChPT) \cite{GL85} is the low-energy effective theory of
Quantum Chromodynamics (QCD).
It is built from the global $SU(3)_L\times SU(3)_R$ symmetry exhibited by QCD in the chiral limit, and described in terms of an octet of pseudoscalar bosons appearing in the theory,
as a result of the spontaneous breaking of this symmetry down to $SU(3)_V$.
The eight Goldstone bosons are identified with the lightest hadronic states $(\pi, K, \eta)$
and their small masses are generated from the quark mass term,
which explicitly breaks the global symmetry of QCD.
The ChPT Lagrangian is organized in terms of an increasing number of powers of momentum and quark masses.
Below the resonance region $(E<M_\rho)$, the interactions of the $(\pi, K, \eta)$ particles are systematically analyzed and easily understood within this framework.
The enormous success in the description of these low-energy interactions makes of ChPT a powerful theoretical tool \cite{Pich:1995bw}.
In this formalism, however, the pseudoscalar singlet $\eta_1$ is not explicitly included
since the $U(1)_A$ anomaly prevents this state from becoming a ninth Goldstone boson.
Therefore, processes involving the $\eta_1$ as an external degree of freedom are not
accounted for in ChPT\footnote{
Strictly speaking, ChPT takes into account the virtual effects of the pseudoscalar singlet
by means of the low-energy constants.
}.
This drawback is remedied in the large-$N_C$ limit, where the effects of the axial anomaly are absent and the symmetry is enlarged to $U(3)_L\times U(3)_R$.
Then, a simultaneous expansion in $p^2$, $m_q$ and $1/N_C$ is possible and
the interactions among the $(\pi, K, \eta, \eta^\prime)$ mesons can be described with a Lagrangian.
Whether this extended framework, named large-$N_C$ ChPT after the work of Kaiser and Leutwyler
\cite{U3-ChPT}, is well established is still under discussion due to the large physical mass of the $\eta^\prime$.
In addition, large-$N_C$ ChPT does not include resonances as external states.
Their masses are supposed to be bigger than the center-of-mass
total energy of a given process involving the pseudoscalar mesons and consequently they are integrated out.
The effects of these resonances are then virtual and encoded in the low-energy constants of the chiral Lagrangian.
However, when the energy of the process is of the order of the resonance mass,
the perturbative expansion of ChPT stops being valid
and the resonant effects must be taken into account explicitly.
This is considered in Resonance Chiral Theory (RChT) \cite{RChTb},
where the interactions of the pseudoscalar mesons are supplemented with new interactions among these and nonets of vectors, axials, scalars, representing the $\rho$, $a_1$, $\sigma$, etc.,
in a minimal way.

The decays $\eta^\prime\to\eta\pi\pi$ are interesting for several reasons.
First, due to the quantum numbers of the pseudoscalar mesons involved,
any resonance involved in the decay must be mainly of scalar nature.
$G$-parity prevents vectors from contributing.
Therefore, this decay is specially suitable for the analysis of the properties of the $f_0(600)$
(or $\sigma$) resonance, even though the $a_0(980)$ is also present and, in fact, dominant.
Second, the presence of $\eta$ and $\eta^\prime$ in this reaction is ideal for studying the mixing properties of these two mesons.
Third, and more general, this decay allows to test ChPT and its possible extensions such as
large-$N_C$ ChPT and RChT.
In view of all that, precision measurements on $\eta$ and $\eta^\prime$ would be very helpful and would provide useful information on our understanding of low-energy QCD.
In particular, their decays are very useful for studying symmetries and symmetry breakings in QCD. The simultaneous treatment of both the $\eta$ and $\eta^\prime$ imposes constraints on theoretical approaches which are tighter than those considering the $\eta$ alone.

Accordingly, there is at present an intense activity studying these processes.
Recently, the GAMS-$4\pi$ and VES Collaborations have measured the related Dalitz plot parameters
(GAMS-$4\pi$ for the $\eta^\prime\to\eta\pi^0\pi^0$ channel \cite{GAMSexp} and
VES for the $\eta^\prime\to\eta\pi^+\pi^-$ one \cite{VESexp})
complementing older results reported by an early GAMS Coll.~\cite{Alde:1986nw} and
CLEO \cite{Briere:1999bp}.
In the isospin limit the values of the Dalitz plot parameters should be the same
but the experimental measurements show some discrepancies among them.
Therefore, there is a clear need for improvement in experimental precision and new measurements of these parameters are foreseen at KLOE-2, Crystal Ball, Crystal Barrel and maybe WASA with improved statistics and a good understanding of the systematical errors.
At KLOE the $\eta^\prime$ is produced via the process $e^+e^-\to\Phi$ followed by
$\Phi\to\eta^\prime\gamma$.
For the $\eta^\prime\to\eta\pi^+\pi^-$ decay channel around 21K events are now on tape.
The background is seen to be very low and these data can easily be used to determine the Dalitz plot parameters.
The increased luminosity at KLOE-2 (a factor 3) will open new possibilities for $\eta^\prime$ studies.
For instance, a Monte Carlo simulation of the $\eta^\prime\to\eta\pi^+\pi^-$
process shows that the detector has a good sensitivity to
the $\sigma$ meson~\cite{Venanzoni:2010ix}.
At Crystal Ball, the $\eta^\prime$ is photo-produced, $\gamma p\to\eta^\prime p$,
and can be identified, \textit{e.g.}, from its decay $\eta^\prime\to\eta\pi^0\pi^0\to 6\gamma$.
About 10K events were expected during 2009 for the $\eta^\prime\to\eta\pi^0\pi^0$ process.
At WASA, large samples of $\eta^\prime$ will be produced in the reaction $pp\to pp\eta^\prime$
as soon as the new detector is developed.
Finally, about 27 million and 12 million decay events could be detected at BES-III each year for
$\eta^\prime\to\eta\pi^+\pi^-$ and $\eta^\prime\to\eta\pi^0\pi^0$, respectively \cite{Li:2009jd}.
On the theory side, the $\eta^\prime\to\eta\pi\pi$ decays have been studied within an effective chiral Lagrangian approach in which the lowest lying scalar mesons are combined into a possible
nonet \cite{Fariborz:1999gr} and, more recently, within the framework of $U(3)$ chiral effective field theory in combination with a relativistic coupled-channels approach \cite{Borasoy:2005du}.
Older analyses based on chiral symmetric frameworks include contact terms
\cite{Cronin:1967jq,Schwinger:1968zz,Di Vecchia:1980sq,Fajfer:1987ij,HerreraSiklody:1999ss}
or contact terms plus scalar meson exchanges \cite{Schechter:1993tc,Singh:1975aq}.

In this work we study the decays $\eta^\prime\to\eta\pi\pi$ in two well established chiral frameworks, large-$N_C$ ChPT and RChT.
Since the charged decay is the most prominent channel of the $\eta^\prime$
we perform the whole analysis for this decay.
Working in the isospin limit, the predictions for the neutral channel will be the same at the level of amplitudes as for the charged channel, and one half at the level of the branching ratios.
Preliminary results were presented in
Refs.~\cite{Masjuanproceeding,Escribanoproceeding}.

The paper is divided as follows.
In Sec.~\ref{DP}, we introduce the Dalitz plot parameterization and discuss the different values of the related parameters.
We also report on the latest experimental values of the charged and neutral branching ratios together with their ratio.
Sec.~\ref{LargeNcChPT} is devoted to the analysis of the $\eta^\prime\to\eta\pi^+\pi^-$ process in the framework of large-$N_C$ ChPT, first computing its predictions at leading and next-to-leading order and then considering the corrections due to the unitarization of the $\pi\pi$ channel.
In Sec.~\ref{RChT}, the decay $\eta^\prime\to\eta\pi^+\pi^-$ is analyzed in the framework of RChT starting with the leading order prediction and later calculating the subleading corrections due to resonance widths and the unitarization of the $\pi\pi$ channel.
Finally, we discuss the results obtained and present our conclusions in Sec.~\ref{conclusions}.
The techniques of partial-wave projection and unitarization of final state interactions (FSI)
used in this work along with a description of $\eta$-$\eta^\prime$ mixing for states and decay constants are included in the appendices for completeness.

\section{Experimental status of Dalitz plot parameters and branching ratios}
\label{DP}
The Dalitz plot distribution for the charged decay channel is described by the following two variables:
\begin{equation}
\label{DPpar}
X=\frac{\sqrt{3}}{Q}(T_{\pi^+}-T_{\pi^-})\ ,\qquad
Y=\frac{m_\eta+2m_\pi}{m_\pi}\frac{T_\eta}{Q}-1\ ,
\end{equation}
where $T_{\pi^\pm ,\eta}$ denote the kinetic energies of mesons in the
$\eta^\prime$ rest frame,
\begin{equation}
T_\eta=\frac{(m_{\eta^\prime}-m_\eta)^2-s}{2 m_{\eta^\prime}}\ ,\quad
T_{\pi^+}=\frac{(m_{\eta^\prime}-m_\pi)^2-t}{2 m_{\eta^\prime}}\ ,\quad
T_{\pi^-}=\frac{(m_{\eta^\prime}-m_\pi)^2-u}{2 m_{\eta^\prime}}\ ,
\end{equation}
and
$Q=T_\eta+T_{\pi^+}+T_{\pi^-}=m_{\eta^\prime}-m_\eta-2m_\pi$.
The Mandelstam variables
$s\equiv (p_{\pi^+}+p_{\pi^-})^2\equiv m_{\pi\pi}^2$,
$t\equiv (p_{\eta^\prime}-p_{\pi^+})^2\equiv m_{\eta\pi}^2$ and
$u\equiv (p_{\eta^\prime}-p_{\pi^-})^2$ have been employed here,
which obey the relation $s+t+u=m_{\eta^\prime}^2+m_\eta^2+2m_\pi^2$.

The squared absolute values of both decay amplitudes (charged and neutral) are expanded around the center of the corresponding Dalitz plot in order to obtain the Dalitz slope
parameters\footnote{
The parameterization in Eq.~(\ref{A2}) had been proposed in Ref.~\cite{Montanet:1994xu}
with an extra term $eXY$.
The analysis of Ref.~\cite{VESexp} included this term in their fits and found that parameter $e$
is consistent with zero.}
\cite{VESexp}:
\begin{equation}
\label{A2}
|A(X,Y)|^2=|{\cal N}|^2[1+aY+bY^2+cX+dX^2]\ ,
\end{equation}
where $a, b, c$ and $d$ are real parameters and $|{\cal N}|^2$ is a normalization factor.
For the charged channel odd terms in $X$ are forbidden due to charge conjugation symmetry,
while for the neutral $c=0$ from symmetry of the wave function.
The Dalitz plot parameters may be different for charged and neutral
decay channels.  However, in the isospin limit they should be the same.
A second parametrization is the linear one \cite{Nakamura:2010zzi}:
\begin{equation}
\label{A2linear}
|A(X,Y)|^2\propto |1+\alpha Y|^2+cX+dX^2\ ,
\end{equation}
where $\alpha$ is a complex parameter.
Comparison with the general fit gives $a=2\mbox{Re}(\alpha)$ and
$b=\mbox{Re}^2(\alpha)+\mbox{Im}^2(\alpha)$.
The two parametrization are equivalent if $b>a^2/4$.

\TABLE{
{\scriptsize
\begin{tabular}{ccccc}
\hline
Parameter &  Exp. $[\eta^\prime\to \eta \pi^0\pi^0]$
&  Th. $[\eta^\prime\to \eta \pi^0\pi^0]$
&  Exp. $[\eta^\prime\to \eta \pi^+\pi^-]$
&  Th. $[\eta^\prime\to \eta \pi^+\pi^-]$
\\
&
GAMS-4$\pi$ \cite{GAMSexp} & Borasoy \& Nissler \cite{Borasoy:2005du} &
VES \cite{VESexp} & Borasoy \& Nissler \cite{Borasoy:2005du} \\
\hline
$a$ & $-0.066\pm 0.016\pm 0.003$ & $-0.127\pm 0.009$
& $-0.127\pm 0.016\pm 0.008$ & $-0.116\pm 0.011$ \\
$b$ & $-0.063\pm 0.028\pm 0.004$ & $-0.049\pm 0.036$
& $-0.106\pm 0.028\pm 0.014$ & $-0.042\pm 0.034$ \\
$c$ & $-0.107\pm 0.096\pm 0.003$ & ---
& $+0.015\pm 0.011\pm 0.014$ &  --- \\
$d$ & $+0.018\pm 0.078\pm 0.006$ & $+0.011\pm 0.021$
& $-0.082\pm 0.017\pm 0.008$ & $+0.010\pm 0.019$ \\
\hline \\
\end{tabular}
\caption{Dalitz slope parameters (experiment and theory) for
$\eta^\prime\to\eta\pi^0\pi^0$ (second and third columns) and
$\eta^\prime\to\eta\pi^+\pi^-$ (fourth and fifth columns), respectively.}
\label{tablepar}
}}

The latest available experimental information on the Dalitz slope parameters is summarized in
Table \ref{tablepar}.
The analysis by the GAMS-4$\pi$ Collaboration is based on approximately 15000 events
\cite{GAMSexp}.
Note that $b$ is negative here and the fit is thus not compatible with the linear fit of
Eq.~(\ref{A2linear}).
If a fit is done with the linear parameterization one gets $\alpha=-0.042\pm 0.008$,
which is in agreement with an early measurement from GAMS based on 5400 events that gave
$\alpha=-0.058\pm 0.013$ (assuming $\mbox{Im}(\alpha)=0$ and $c=0$) \cite{Alde:1986nw}.
Both analyses are for the $\eta^\prime\to\eta\pi^0\pi^0$ channel.
Regarding the $\eta^\prime\to\eta\pi^+\pi^-$ decay, the result from CLEO based on 6700 events yields $\alpha=-0.021\pm 0.025$ (assuming $\mbox{Im}(\alpha)=0$, $c=0$ and $d=0$)
using the same linear fit as for the neutral decay channel \cite{Briere:1999bp}.
The VES analysis is based on roughly 13600 events obtained from the charge-exchange reaction
$\pi^-p\to\eta^\prime n$ and 6500 events from the diffractive-like production reaction
$\pi^-N\to\eta^\prime\pi^- N$ \cite{VESexp}.
A fit using the combined data sets from VES gave the parameter values shown in
Table \ref{tablepar}.
Again, $b$ is found to be negative and thus incompatible with a linear fit,
while $c$ is consistent with zero.
The previous work supersedes a first study \cite{Amelin:2005rz},
where the $\eta\pi^+\pi^-$ Dalitz plot has been investigated with a sample of approximately
7000 events obtaining $\alpha=-0.072\pm 0.012\pm 0.006$
---this is the real part of $\alpha$ while $\mbox{Im}(\alpha)=0.0\pm 0.1\pm 0.0$---
in the linear and $a=-0.120\pm 0.027\pm 0.015$ in the general parameterization.
The $C$-violation parameter was compatible with zero, $c=0.021\pm 0.024$.
In average, $\alpha=-0.059\pm 0.011$ \cite{Nakamura:2010zzi}.
Finally, a measurement of $\alpha$ not included in the average is reported in
Ref.~\cite{Kalbfleisch:1974ku}, where $\alpha=-0.08\pm 0.03$
(assuming $\mbox{Im}(\alpha)=0$ and $c=0$) with about 1400 events.

For the general parameterization, we see from Table \ref{tablepar} that there is some tension in all  the parameters.
Notice, however, that the theory model results of Ref.~\cite{Borasoy:2005du}
for the charged and neutral decay channels are compatible among themselves.
While there is agreement in $a$ between the VES fitted value and the result of
Ref.~\cite{Borasoy:2005du} there is not such when compared to the GAMS-$4\pi$ value.
The same happens to $d$ but this time the GAMS-$4\pi$ and the theory model results agree,
although they are in conflict with the VES reported value.
The case of $b$ is less severe.
The VES value is compatible with the others only at 2$\sigma$.
Finally, there is a variation between the measured values of $c$.
Nevertheless, the large statistical error of the GAMS-$4\pi$ result makes this statement not conclusive.

In this work we want to point out the possibility and maybe the need of extending the
Dalitz plot parameterization up to higher orders in a systematic way.
Hence, we perform a general power expansion in
$Y$ and $X^2$ (odd powers of $X$ are forbidden)
and consider the following extended parameterization
\begin{equation}
|A(X,Y)|^2=|\mN|^2
[1+(aY+dX^2)+(bY^2+\kappa_{21}X^2 Y+\kappa_{40} X^4)+\cdots\ .
\label{eq.slope-par2}
\end{equation}
Higher terms of the form  $\kappa_{2m,n}X^{2m} Y^n$ with $m+n\geq 3$ are not considered,
since they are beyond the precision of current and forthcoming experiments.
The parameters $a$, $b$ and $d$ have been left with the original nomenclature
in order to better compare to former analyses.

Finally, we present the latest experimental values for the charged and neutral
$\eta^\prime\to\eta\pi\pi$ branching ratios.
The CLEO collaboration has recently measured both channels from the analysis of
$J/\psi\to\eta^\prime\gamma$ events \cite{:2009tia} and obtained
$\mB_{\eta^\prime\to\eta\pi^+\pi^-}=0.424\pm 0.011\pm 0.004$ and
$\mB_{\eta^\prime\to\eta\pi^0\pi^0}=0.235\pm 0.013\pm 0.004$, respectively.
However, we prefer to use the fitted values appearing in the most recent version of the
Review of Particle Properties \cite{Nakamura:2010zzi},
that is, $\mB_{\eta^\prime\to\eta\pi^+\pi^-}=0.432\pm 0.007$ and
$\mB_{\eta^\prime\to\eta\pi^0\pi^0}=0.217\pm 0.008$.
Using these values one gets for the ratio of charged to neutral decay widths,
\begin{equation}
\label{ratio}
\frac{\Gamma(\eta^\prime\to\eta\pi^+\pi^-)}{\Gamma(\eta^\prime\to\eta\pi^0\pi^0)}
=1.99\pm 0.08\ ,
\end{equation}
in perfect agreement with isospin symmetry conservation.

\section{Large-\mbox{\boldmath $N_C$} Chiral Perturbation Theory prediction}
\label{LargeNcChPT}
\subsection{Framework and lowest order calculation}
Large-$N_C$ Chiral Perturbation Theory is the effective field theory of QCD in the chiral and large-$N_C$ limits \cite{U3-ChPT}.
In the large-$N_C$ limit the $U(1)_A$ anomaly is absent and the pseudoscalar singlet  $\eta_1$ becomes the ninth Goldstone boson associated with the spontaneous breaking of
$U(3)_L\times U(3)_R$ down to $U(3)_V$.
Both chiral and large-$N_C$ corrections are treated perturbatively.
The effective Lagrangian is thus organized as a simultaneous expansion in powers of momenta, quark masses and $1/N_C$ \cite{U3-ChPT,LeutwylerBounds},
\begin{equation}
\label{Leff}
{\cal L}_{\rm eff}={\cal L}^{(0)}+{\cal L}^{(1)}+{\cal L}^{(2)}+\cdots\ ,
\end{equation}
where the contributions of order $1$, $\delta$, $\delta^2$, $\ldots$
follow the ordering of the series with
\begin{equation}
\label{powercounting}
\partial_{\mu}={\cal O}(\sqrt{\delta})\ ,\quad m_q={\cal O}(\delta)\ ,\quad 1/N_C={\cal O}(\delta)\ .
\end{equation}
At lowest order, the ${\cal L}^{(0)} = {\cal O}(\delta^0)$ has the form\footnote{
For the purpose of our work external sources are not required.
The Lagrangian including them can be found in Ref.~\cite{U3-ChPT}.}
($\langle A\rangle$ stands for the trace of $A$)
\begin{equation}
\label{L0}
\begin{array}{l}
{\cal L}^{(0)}=
{f^2\over 4}\langle\partial_\mu U^\dagger\partial^\mu U\rangle +
{f^2\over 4}\langle U^\dagger \chi + \chi^\dagger U \rangle -
\frac{1}{2}M_0^2\eta_1^2\ ,
\quad
\end{array}
\end{equation}
where $\chi=2B_0{\cal M}$, $B_0\sim{\cal O}(N^0_C)$ is related to the quark condensate,
${\cal M}=\mbox{diag}(m_u,m_d,m_s)$ is the quark-mass matrix,
$f \sim{\cal O}(\sqrt{N_C})$ is the pion decay constant in the chiral limit,
$M_0^2$ is the $U(1)_A$ anomaly contribution to the $\eta_1$ mass,
and $U=u^2=\exp{(i\sqrt{2}\Phi/f)}$ with
\begin{equation}
\label{goldstons}
\Phi=\left(
\begin{array}{ccc}
\frac{1}{\sqrt{2}}\pi^0 + \frac{1}{\sqrt{6}} \eta_8 + \frac{1}{\sqrt{3}}\eta_1 & \pi^+ & K^+\\[1ex]
\pi^- & -\frac{1}{\sqrt{2}}\pi^0 +\frac{1}{\sqrt{6}}\eta_8 + \frac{1}{\sqrt{3}}\eta_1& K^0\\[1ex]
K^- & \bar{K}^0 & -\frac{2}{\sqrt{6}}\eta_8 + \frac{1}{\sqrt{3}}\eta_1
\end{array}
\right)\ .
\end{equation}
In this case, the mathematical states $\eta_B^T\equiv (\eta_8,\eta_1)$ are related to the physical states $\eta_P^T\equiv (\eta,\eta^\prime)$ by
\begin{equation}
\label{LOmixing}
\left(
\begin{array}{c}
\eta_8\\
\eta_1
\end{array}
\right)=
\left(
\begin{array}{cc}
\cos\theta_P & \sin\theta_P\\
-\sin\theta_P  & \cos\theta_P
\end{array}
\right)
\left(
\begin{array}{c}
\eta\\
\eta^\prime
\end{array}
\right)\ ,
\end{equation}
where $\theta_P$ is the $\eta$-$\eta^\prime$ mixing angle in the octet-singlet basis at this order.
At next-to-leading order, the part of ${\cal L}^{(1)} = {\cal O}(\delta)$ relevant to the analysis of
$\eta^\prime\to\eta\pi\pi$ is
\begin{eqnarray}
\label{L1Nc}
{\cal L}^{(1)}_{\eta^\prime\to\eta\pi\pi}&=&
L_2\langle\partial_\mu U^\dagger\partial_\nu U\partial^\mu U^\dagger \partial^\nu U\rangle
+(2 L_2 + L_3)\langle\partial_\mu U^\dagger\partial^\mu U\partial_\nu U^\dagger\partial^\nu U\rangle\nonumber\\[1ex]
&&+ L_5\langle\partial_\mu U^\dagger\partial^\mu U (U^\dagger\chi+\chi^\dagger U \big)
\rangle
+L_8\langle U^\dagger\chi U^\dagger\chi + \chi^\dagger U \chi^\dagger U
\rangle\nonumber\\[1ex]
&&+ {\Lambda_1\over 2}\partial_\mu\eta_1\partial^\mu\eta_1
      -i{f\Lambda_2\over 2\sqrt{6}}\eta_1\langle U^\dagger\chi -  \chi^\dagger U\rangle\ ,
\end{eqnarray}
where the low-energy constants (LECs) $L_2, L_3, L_5, L_8$ are of ${\cal O}(N_C)$,
while $\Lambda_1, \Lambda_2$ are of ${\cal O}(1/N_C)$.
$\Lambda_1$ and $\Lambda_2$ only influence the singlet sector and can be attributed to OZI-rule violating contributions.
Now, the relation between the mathematical and physical states cannot be simply expressed in terms of a mixing angle.
The precise connection is found in Eq.~(\ref{NLOmixing}).

At lowest order (LO), \textit{i.e.}~using the ${\cal L}^{(0)}$ in Eq.~(\ref{L0}),
the amplitude is
\cite{Cronin:1967jq,Schwinger:1968zz,Di Vecchia:1980sq,HerreraSiklody:1999ss,
Fajfer:1987ij,Schechter:1993tc,Singh:1975aq,BijnensEtas}
\begin{eqnarray}
\label{amplLO}
{\cal M}_{\eta^\prime\to\eta\pi\pi}^{\rm ChPT}|_{\rm LO}=
\left[2\sqrt{2}\cos(2\theta_P)-\sin(2\theta_P)\right]{m^2_{\pi} \over 6f^2}\ ,
\end{eqnarray}
where $f=f_\pi$ (the pion decay constant) at this order.
For $\theta_P=(-13.3\pm 0.5)^\circ$ \cite{Ambrosino:2009sc} and $f_\pi=92.2$ MeV,
the branching ratio thus obtained is only 3\% of the measured value.
The main reason for this difference is the unexpected appearance of $m_{\pi}^2$ in the amplitude
(\ref{amplLO}) that makes it to vanish in the chiral limit.
Although in general the LO is able to provide a suitable approximation
to the hadronic amplitude,
\textit{e.g.}~the $\eta$-$\eta^\prime$ mixing \cite{HerreraSiklody:1999ss,LKdecays}
or the $\eta(\eta^\prime)\to\gamma \gamma$ decay
\cite{HerreraSiklody:1999ss,BijnensEtas,Bijnens-anomalous,Siklodyetaetapmixing,EFTPich},
in some few cases, like the anomalous magnetic moment $g-2$ \cite{Bijnensg2}
or the $\gamma \gamma\to\pi^0\pi^0$ scattering \cite{Gasser-ggtopipi},
the LO contribution is absent (or very small) and the first (and dominant) contribution comes from higher orders.

\subsection{Next-to-leading order calculation}
\label{NLOChPT}

At next-to-leading order (NLO),
it is convenient to express the amplitude, from the point of view of large-$N_C$,
in terms of OZI-allowed (${\cal M}_{\eta_q \eta_q \pi\pi}$)
and OZI-suppressed (${\cal M}_{\eta_s \eta_q \pi\pi}$ and ${\cal M}_{\eta_s \eta_s \pi\pi}$) contributions\footnote{
The octet-singlet states $(\eta_8,\eta_1)$ are related to
their quark-flavour counterparts $(\eta_q,\eta_s)$ by
$(\eta_8,\eta_1)=(\eta_q-\sqrt{2}\eta_s,\sqrt{2}\eta_q+\eta_s)/\sqrt{3}$.},
\begin{equation}
{\cal M}_{\eta^\prime\to\eta\pi\pi}^{\rm ChPT}|_{\rm NLO}=
c_{qq}{\cal M}_{\eta_q \eta_q \pi\pi}+
c_{sq}{\cal M}_{\eta_s \eta_q \pi\pi}+
c_{ss}{\cal M}_{\eta_s \eta_s \pi\pi}\ ,
\end{equation}
where the coefficients $c_{qq}$, $c_{sq}$ and $c_{ss}$
($q$ stands for the $u$ and $d$ quarks and $s$ for the $s$ quark) are universal
and encode the $\eta$-$\eta^\prime$ mixing at next-to-leading order
(see App.~\ref{app.eta-mix} for details),
\begin{eqnarray}
c_{qq}&=&
-\frac{f^2}{3 f_8^2 f_0^2 \cos^2(\theta_8-\theta_0)}
\left[2f_8^2\sin(2\theta_8) - f_0^2 \sin(2\theta_0) - 2\sqrt{2} f_8 f_0\cos(\theta_8+\theta_0)\right]\ ,
\nonumber\\
c_{sq} &=&
-\frac{f^2}{3 f_8^2 f_0^2 \cos^2(\theta_8-\theta_0)}
\left[\sqrt{2} f_8^2\sin(2\theta_8) + \sqrt{2} f_0^2\sin(2\theta_0) +
f_8 f_0\cos(\theta_8+\theta_0)\right]\ ,
\nonumber\\
c_{ss} &=&
-\frac{f^2}{3 f_8^2 f_0^2 \cos^2(\theta_8-\theta_0)}
\left[f_8^2\sin(2\theta_8) - 2 f_0^2\sin(2\theta_0) + 2 \sqrt{2} f_8 f_0\cos(\theta_8+\theta_0)\right]\ ,
\label{eq.cqq-mixing}
\end{eqnarray}
and
\begin{eqnarray}
{\cal M}_{\eta_q \eta_q \pi\pi}&=&
{1\over f^2}\left[
\frac{2(3L_2+ L_3)}{f_\pi^2}\left(s^2+t^2+u^2-m^4_{\eta^\prime}-m^4_{\eta}-2 m^4_{\pi}\right)
\right.\nonumber\\
&&
\left.
-\frac{2 L_5}{f_\pi^2}\left(m^2_{\eta^\prime}+m^2_{\eta}+2 m^2_{\pi}\right)m^2_{\pi}
+\frac{24 L_8}{f_\pi^2}m^4_{\pi}+\frac{2}{3}\Lambda_2 m^2_{\pi}\right]\ ,\\[1ex]
{\cal M}_{\eta_s \eta_q \pi\pi}&=&
\frac{\sqrt{2}}{3 f^2}\Lambda_2 m^2_{\pi}\ ,\quad
{\cal M}_{\eta_s \eta_s \pi\pi}=0\ ,
\end{eqnarray}
where these amplitudes are independent of the way $\eta$ and $\eta^\prime$ do mix.
Taking into account both the contributions of lowest and next-to-leading order,
the amplitude for $\eta^\prime\to\eta\pi\pi$ is finally given by
\begin{eqnarray}
{\cal M}_{\eta^\prime\to\eta\pi\pi}^{\rm ChPT}&=&
{c_{qq}\over f^2}\left[\frac{m_\pi^2}{2}+
\frac{2(3L_2+ L_3)}{f_\pi^2}\left(s^2+t^2+u^2-m^4_{\eta^\prime}-m^4_{\eta}-2 m^4_{\pi}\right)
\right.\nonumber\\
&&
\left.
-\frac{2 L_5}{f_\pi^2}\left(m^2_{\eta^\prime}+m^2_{\eta}+2 m^2_{\pi}\right)m^2_{\pi}
+\frac{24 L_8}{f_\pi^2}m^4_{\pi}+\frac{2}{3}\Lambda_2 m^2_{\pi}\right]\nonumber\\[1ex]
&&
+\frac{c_{sq}}{f^2}\frac{\sqrt{2}}{3}\Lambda_2 m^2_{\pi}\ .
\label{amplLN}
\end{eqnarray}
As seen, the only contribution not proportional to $m_\pi^2$ and therefore dominant is the $3L_2+L_3$ term.
Indeed, this term is larger than the LO prediction which in turn is larger than the sum of the NLO
$L_5, L_8$ and $\Lambda_2$ terms.
Consequently, the branching ratio is expected to be of the form
$(\tilde{a} + \tilde{b}\, m_\pi^2 + \tilde{c}\, m_\pi^4)^2
\simeq \tilde{a}^2 + 2\,\tilde{a}\,\tilde{b}\, m_\pi^2$,
where $\tilde{a}$ corresponds to the $3L_2+L_3$ term
and $\tilde{b}$ and $\tilde{c}$ are the suppressed contributions.

For the numerical analysis,
we use $f_8=1.28f_{\pi}$, $f_0\simeq 1.25 f_{\pi}$,
$\theta_8\simeq -20^\circ$, $\theta_0\simeq -4^\circ$
and $\Lambda_2\simeq 0.3$ from Ref.~\cite{Leutwyler:1997yr},
and $m_{\pi}=137.3$ MeV, $m_{\eta}=547.9$ MeV and $m_{\eta^\prime}=957.8$ MeV from
Ref.~\cite{Nakamura:2010zzi}.
For the LECs in the large-$N_C$ limit we find in the literature three different sets of values gathered in Table~\ref{tab.LECs}.
All of them come from resonance-exchange
estimations with only one resonance per channel~\cite{ChPTEcker,BEGLis,PichLis}.
In the simplest scheme one obtains the set of LECs identified as \textit{set1} \cite{ChPTEcker}.
The values of the LECs are slightly modified after including QCD-inspired assumptions of
high-energy behaviour\footnote{
Using a model with a finite number of resonances instead of an infinite one,
which would correspond to the full large-$N_C$ theory,
may introduce problems in the short-distance matching and large discrepancies in the
prediction of LECs \cite{PAs}.}.
When an unsubtracted dispersion relation for the pion vector form factor is used
one obtains \textit{set2} \cite{BEGLis}.
When constraints on the scalar form factor are also used one gets \textit{set3} \cite{PichLis}.

\TABLE{
\begin{tabular}{|c|r|r|r|}
  \hline
   & \textit{set1} \cite{ChPTEcker} & \textit{set2} \cite{BEGLis} & \textit{set3} \cite{PichLis} \\
   \hline
  $L_2\cdot 10^3$ & 1.2 & 1.8 & 1.8 \\
  $L_3\cdot 10^3$ & $-3.0$ & $-4.9$ & $-4.3$ \\
  $(3L_2+L_3)\cdot 10^3$ & 0.6 & 0.5 & 1.1 \\
  $L_5\cdot 10^3$ & 1.4 & 1.4 & 2.1 \\
  $L_8\cdot 10^3$ & 0.9 & 0.9 & 0.8 \\
  \hline
\end{tabular}
\caption{Values for the three different sets of low-energy constants discussed in the text.}
\label{tab.LECs}
}

The predicted branching ratio for the different sets of LECs is found to be
14\% \textit{(set1)}, 9\% \textit{(set2)} and 64\% \textit{(set3)}.
As seen, the next-to-leading results are in general one order of magnitude bigger than the lowest order prediction after Eq.~(\ref{amplLO}), thus indicating that our approach for the description of
$\eta^\prime\to\eta\pi\pi$ moves into the right direction.
However, due to the broad range of values obtained for the predicted branching ratio, caused by the numerical difference in the $3L_2+L_3$ combination,
it seems to be more appropriate to fix the value of this dominant contribution from the measured branching ratio while keeping the remaining LECs to the values shown in Table \ref{tab.LECs}
for the corresponding set.
Since $3L_2+L_3$, $L_5$ and $L_8$ are similar in \textit{set1} and \textit{set2},
and different with respect to \textit{set3},
we will use for the purpose of comparison only \textit{set1} (without QCD-inspired constraints) and
\textit{set3} (with QCD-inspired constraints).
In Table \ref{tab.dalitzparamLNcfit}, the outcome for the $3L_2+L_3$ combination obtained in this way are displayed.
The values seem to be in line with \textit{set3}.

Besides the branching ratio, the study of the Dalitz plot distribution provides relevant information.
The squared amplitude $|{\cal M}|^2$ is expanded in terms of the variables $X$ and $Y$
(see Sec.~\ref{DP} for details), where $Y$ is a linear function of $s$, the $\pi\pi$ invariant mass,
and $X$ appears always in the form of $\cos\theta_\pi=X f(Y)$, with $\theta_\pi$ the angle of
$\mathbf{p}_{\pi^+}$ with respect to $\mathbf{p}_{\eta}$ in the $\pi\pi$ rest frame.
This explains the reason for the absence of odd powers of $X$ under the assumption of $C$-invariance.
Interesting facts concerning the Dalitz plot parameters are the following.
First, the amplitude (\ref{amplLN}) in the chiral limit (with $m_\eta, m_{\eta^\prime}\neq 0$) becomes
\begin{equation}
\label{amplmpi0LN}
\lim_{m_\pi\to 0}{\cal M}_{\eta^\prime\to\eta\pi\pi}^{\rm ChPT}=
-\frac{c_{qq}}{f^2}\frac{4(3L_2+L_3)}{3f_\pi^2}
m_{\eta^\prime}(m_{\eta^\prime}-m_{\eta})^2 (6 m_{\eta}-m_{\eta^\prime} X^2)\ .
\end{equation}
This means that when ${\cal M}$ is expanded in $X$ and $Y$ around the center of the Dalitz plot the terms with powers of $Y$ will be proportional to $m_\pi$ and because of that suppressed.
As a result, we find that the terms in the expansion of order $Y$ are similar in magnitude to those of order $X^2$.
Then, one would expect for the Dalitz plot parameters the following hierarchy:
$a\sim d\gg b\sim\kappa_{21}\sim\kappa_{40}$.
For this reason, we propose to extend the standard Dalitz plot parameterization used by experimental analyses (based on $a$, $b$ and $d$ alone) and include the additional
$\kappa_{21}$ and $\kappa_{40}$ parameters for consistency.
Second, certain combinations of parameters turn out to be independent of the chiral couplings,
being just functions of the pseudoscalar masses.
In particular, we obtain
\begin{equation}
\label{dalitzparamrelations}
\frac{a}{d}=\frac{\kappa_{21}}{2\kappa_{40}}=
-\frac{6m_{\pi}(m_{\eta}-m_{\pi})(m_{\eta^\prime}^2-(m_{\eta}+2m_{\pi})
(3m_{\eta^\prime}-m_{\eta}-m_{\pi}))}
{m_{\eta^\prime}(m_{\eta}+2m_{\pi})^2(m_{\eta^\prime}-m_{\eta}-2m_{\pi})}=3.4\ .
\end{equation}
Our predictions for the Dalitz plot parameters are summarized in Table \ref{tab.dalitzparamLNcfit},
together with the reported values from GAMS-$4\pi$ \cite{GAMSexp} and VES \cite{VESexp} based on analyses of $\eta^\prime\to\eta\pi^0\pi^0$ and $\eta^\prime\to\eta\pi^+\pi^-$, respectively.
The parameters are almost identical for both sets thanks to the dominance of the $3 L_2+L_3$ combination over the tiny contributions from $L_5$, $L_8$ and $\Lambda_2$ terms.
When compared to the measured values the $d$ parameter is in the correct range
whereas the $a$ and $b$ are clearly incompatible.
This points out the need for including higher order effects.
Final state interactions originated from the rescattering of the two pions are the main source of such effects.

\TABLE{
\begin{tabular}{|l|c|c|c|c|}
  \hline
   Parameter & \textit{set1}	& \textit{set3}	&
   GAMS-4$\pi$	 \cite{GAMSexp}	& VES  \cite{VESexp}\\
   \hline
   $a[Y]$					& $-0.303$	& $-0.284$	& $-0.066\pm 0.016$
   									& $-0.127\pm 0.018$\\
   $b[Y^2]$				& $+0.001$   	& $-0.001$	& $-0.063\pm 0.028$
   									& $-0.106\pm 0.031$\\
   $d[X^2]$				& $-0.089$	& $-0.084$	& $+0.018\pm 0.078$
   									& $-0.082\pm 0.019$\\
   $\kappa_{21}[X^2Y]$		& $+0.014	$	& $+0.012$	& ---		& ---		\\
   $\kappa_{40}[X^4]$		& $+0.002$	& $+0.002$	& ---		& ---		\\
   \hline
   $(3L_2+L_3)\cdot 10^3$	& 1.0			& 0.9			&		&		\\
   \hline
\end{tabular}
\caption{Dalitz plot parameters for \textit{set1} and \textit{set3}
and the measured values (errors are added in quadrature)
from $\eta^\prime\to\eta\pi^0\pi^0$ (GAMS-4$\pi$) and $\eta^\prime\to\eta\pi^+\pi^-$ (VES).
The combination $3L_2+L_3$ is extracted for each set from the experimental branching ratio.}
\label{tab.dalitzparamLNcfit}
}

\subsection{$\pi\pi$ final state interactions}

As stated, a further improvement in the large-$N_C$ approach would be to calculate
the next-to-next-to-leading (NNLO) effects.
In this case, loops come into play and the LECs are scale-dependent.
A detailed calculation of the loop effects would account for the related final state interactions.
In view of the exploratory nature of our analysis this calculation is beyond the scope of the present work.
However, we can estimate these rescattering effects by means of a unitarization procedure.
Different from scattering processes, in $\eta^\prime\to\eta\pi\pi$ all the channels contribute to the unitarity relation (see App.~\ref{unitarization} for details).
Nonetheless, the contributions from $\eta\pi$ rescattering effects ($t$- and $u$-channel) are negligible.
This was also remarked in a previous non-relativistic effective field theory study~\cite{Kubis1},
where they found a very small value for the $\eta\pi$ scattering length.
Accordingly, we only consider $s$-channel unitarity due to $\pi\pi$ final state interactions.
This unitarity constraint is better expressed in terms of partial waves (see Eq.~(\ref{eq.unitarity})).
The constraint incorporates the $\pi\pi$ $I=0$ partial-wave amplitudes ${\cal T}^0_J$,
which for $J=0,2$ are found in App.~\ref{pwp}.

For $\eta^\prime\to\eta\pi\pi$ the relevant partial waves are $J=0,2$ (higher partial waves are suppressed by phase space) since $I(\pi\pi)=0$ and $I+J=0,2\ldots$ must be fulfilled.
Hence, we compute the $S$- and $D$-wave projections of the amplitude (\ref{amplLN})
as specified by Eq.~(\ref{partialwave}).
The result is
\begin{eqnarray}
{\cal M}_0(s)  &=&
\frac{1}{192\pi}
\left\{
\frac{c_{qq}}{f^2}
\left[3 m_{\pi}^2
+\frac{4 (3L_2+L_3)}{f_\pi^2}
\left(
(s-m_{\eta^\prime}^2)(5 s+(m_{\eta^\prime}^2-8 m_{\pi}^2)+2 m_{\eta^\prime}^2 m_{\eta}^2/s)
\right.\right.\right.\nonumber\\[1ex]
&&\left.\left.\left.
- 4 s m_{\eta}^2+m_{\eta}^2 (2 m_{\eta^\prime}^2 - m_{\eta}^2 + 10 m_{\pi}^2)
+ 2 m_{\eta}^2 m_{\pi}^2 (2 m_{\eta^\prime}^2 - m_{\eta}^2)/s
\right)
\right.\right.\nonumber\\[1ex]
&&\left.\left.
-12 m_{\pi}^2
\left(\frac{L_5}{f_\pi^2}(m_{\eta^\prime}^2+m_{\eta}^2+2 m_{\pi}^2)
-\frac{12L_8}{f_\pi^2}m_{\pi}^2\right)
+4\Lambda_2 m_{\pi}^2
\right]
\right.\nonumber\\[1ex]
&&\left.
+\frac{c_{sq}}{f^2}2\sqrt{2}\Lambda_2 m_{\pi}^2
\right\}\ ,\\[2ex]
{\cal M}_2(s)  &=&
\frac{1}{240\pi}\frac{c_{qq}}{f^2}\frac{3L_2+L_3}{f_\pi^2}(s-4 m_{\pi}^2)
\left(s-2(m_{\eta^\prime}^2 + m_{\eta}^2) + (m_{\eta^\prime}^2 - m_{\eta}^2)^2/s\right)\ .
\label{eq.MJ-ChPT}
\end{eqnarray}
The predicted branching ratio is completely dominated by the $S$-wave (scalar component),
while the Dalitz plot parameters are not.
The reason is that the variable $X$ always appears through $\cos\theta_\pi$ and,
as a consequence, if only the $S$-wave is considered the terms proportional to $X$ are absent and $d[X^2]=\kappa_{21}[X^2 Y]=\kappa_{40}[X^4]=0$.
The first contribution to the former three parameters comes from the $D$-wave,
which turns out to be essential for their precise determination.
For the $a$ and $b$ parameters the contribution of the $D$-wave is clearly less important than the  scalar component.

Once the partial wave amplitudes are known, one can proceed to unitarize them.
We use two different procedures (see App.~\ref{unitarization} for details),
one based on the $K$-matrix formalism in Eq.~(\ref{amplkmatrix})
and the other inspired by the $N/D$ method in Eq.~(\ref{amplB0}).
The final $\eta^\prime\to\eta\pi\pi$ unitarized invariant amplitudes are
\begin{equation}
\label{amplkmatrixunit}
{\cal M}(s,t,u)|_{\rm K-matrix}=
\sum_J 32\pi (2J+1) P_J(\cos\theta_\pi)
\frac{{\cal M}_J(s)|_{\rm tree}}{1-i\rho(s){\cal T}^0_J(s)|_{\rm tree}}\ ,
\end{equation}
in the simplest $K$-matrix formalism, and
\begin{equation}
\label{amplB0unit}
{\cal M}(s,t,u)|_{\rm N/D}=
\sum_J 32\pi (2J+1) P_J(\cos\theta_\pi)
\frac{{\cal M}_J(s)|_{\rm tree}}{1-16\pi B_0(s){\cal T}^0_J(s)|_{\rm tree}}\ ,
\end{equation}
in the $N/D$ method.

\TABLE{
\begin{tabular}{|l|c|c|c|c|}
  \hline
   Parameter & \textit{set1}	& \textit{set3}	&
   GAMS-4$\pi$	 \cite{GAMSexp}	& VES  \cite{VESexp}\\
   \hline
   $a[Y]$					& $-0.267$	& $-0.248$	& $-0.066\pm 0.016$
   									& $-0.127\pm 0.018$\\
   $b[Y^2]$				& $-0.018$   	& $-0.020$	& $-0.063\pm 0.028$
   									& $-0.106\pm 0.031$\\
   $d[X^2]$				& $-0.089$	& $-0.084$	& $+0.018\pm 0.078$
   									& $-0.082\pm 0.019$\\
   $\kappa_{21}[X^2Y]$		& $+0.010$	& $+0.009$	& ---		& ---		\\
   $\kappa_{40}[X^4]$		& $+0.002$	& $+0.002$	& ---		& ---		\\
   \hline
   $(3L_2+L_3)\cdot 10^3$	& 1.0			& 0.9			&		&		\\
   \hline
\end{tabular}
\caption{Dalitz plot parameters obtained from the amplitude in Eq.~(\ref{amplkmatrixunit})
using the $K$-matrix formalism.
The experimental values are also shown for comparison.}
\label{ttab.Kmdalitzparam}
}

Once these amplitudes are computed, the corresponding Dalitz plot parameters can be extracted.
The results within the $K$-matrix unitarization procedure are shown in
Table \ref{ttab.Kmdalitzparam}.
The combination $3L_2+L3$ is again fixed for each set of LECs from the experimental branching ratio.
Notice that when compared with the results in Table \ref{tab.dalitzparamLNcfit},
the parameters only carrying powers of $X$, that is $d$ and $\kappa_{40}$,
do not change, thus indicating that the unitarized version of the $D$-wave does not produce any quantitative modification.
The reason is that the ${\cal T}^0_2$ $\pi\pi$-scattering amplitude entering into the unitarization of the $D$-wave emerges at next-to-leading order in the chiral expansion.
On the contrary, the parameters carrying powers of $Y$ are substantially modified because of the unitarization of the $S$-wave
(the main contribution to ${\cal T}^0_0$ is in this case the leading order).
When compared to the measured values, the parameters $a[Y]$ and $b[Y^2]$ have changed in the right direction, but not enough, and $d[X^2]$ still agrees.

For the unitarization inspired by the $N/D$ method
(a sophisticated version of the $K$-matrix  formalism),
the constant $C$ involved in the $B_0(s,m_\pi^2,m_\pi^2)$ integral of Eq.~(\ref{amplB0unit}),
see also Eq.~(\ref{eq.B0-def}),
must be fixed from experiment together with the combination $3L_2+L_3$.
In order to fix $C$ we use the better measured Dalitz plot parameter, $a[Y]$,
which is either $-0.066(16)$ \cite{GAMSexp} or $-0.127(18)$ \cite{VESexp}.
This means that the parameter ranges $-0.145<a<-0.050$,
or equivalently $a=-0.098(48)$, in accordance with both experiments.
We keep using the experimental branching ratio to fix $3L_2+L_3$.
The results are shown in Table \ref{tab.Bunidalitzparam}.
The parameters $b$ and $d$ are now in line with the measured values.

\TABLE{
{\small
\begin{tabular}{|l|c|c|c|c|}
  \hline
   Parameter & \textit{set1}	& \textit{set3}	&
   GAMS-4$\pi$	 \cite{GAMSexp}	& VES  \cite{VESexp}\\
   \hline
   $b[Y^2]$				& $-0.051(1)$   	& $-0.050(1)$	& $-0.063\pm 0.028$
   									& $-0.106\pm 0.031$\\
   $d[X^2]$				& $-0.101(8)$	& $-0.092(8)$	& $+0.018\pm 0.078$
   									& $-0.082\pm 0.019$\\
   $\kappa_{21}[X^2Y]$		& $+0.004(2)$	& $+0.003(2)$	& ---		& ---		\\
   $\kappa_{40}[X^4]$		& $+0.003(1)$	& $+0.002(1)$	& ---		& ---		\\
   \hline
   $(3L_2+L_3)\cdot 10^3$	& 1.1(1)		& 1.0(1)		&		&		\\
   $C$-constant 			& $-2.0 \leq C \leq 0.4$
   						& $-1.4 \leq C \leq 0.9$		&		&		\\
   \hline
\end{tabular}
\caption{Dalitz plot parameters obtained from the amplitude in Eq.~(\ref{amplB0unit})
inspired by the $N/D$ method.
The range of the constant $C$ is fixed from $a[Y]=-0.098(48)$.
The errors account for the propagation of the uncertainty in $a[Y]$.}
\label{tab.Bunidalitzparam}
}}

To sum up, rescattering effects are seen to be important in $\eta^\prime\to\eta\pi\pi$.
The unitarization procedure give us an idea of what is missing when the amplitude is computed only up to NLO in the large-$N_C$ ChPT expansion.
From the results in Table \ref{tab.dalitzparamLNcfit} to those in Table \ref{tab.Bunidalitzparam},
there is a clear improvement on the predictions for the Dalitz plot parameters
as compared to the measured values.
Finally, the different parameters agree well within experimental errors.
The next logical refinement would be to incorporate the lowest lying set of resonances and
check wether they are the sole responsible for such improvement.
We postpone this analysis to the next section.

For the benefit of future experimental analyses,
we display in Fig.~\ref{fig.LNChPT-Dalitz},
the Dalitz plot distribution of $\eta^\prime\to\eta\pi\pi$ using Eq.~(\ref{amplLN}),
and in Fig.~\ref{Fig.InvMassLNChPT},
the corresponding invariant mass spectra from
Eq.~(\ref{amplLN}) ---blue dashed line--- and Eq.~(\ref{amplB0unit}) ---solid orange band.
In both figures, we use \textit{set3} for the LECs.

\FIGURE{
\includegraphics[angle=0,clip,width=7cm]{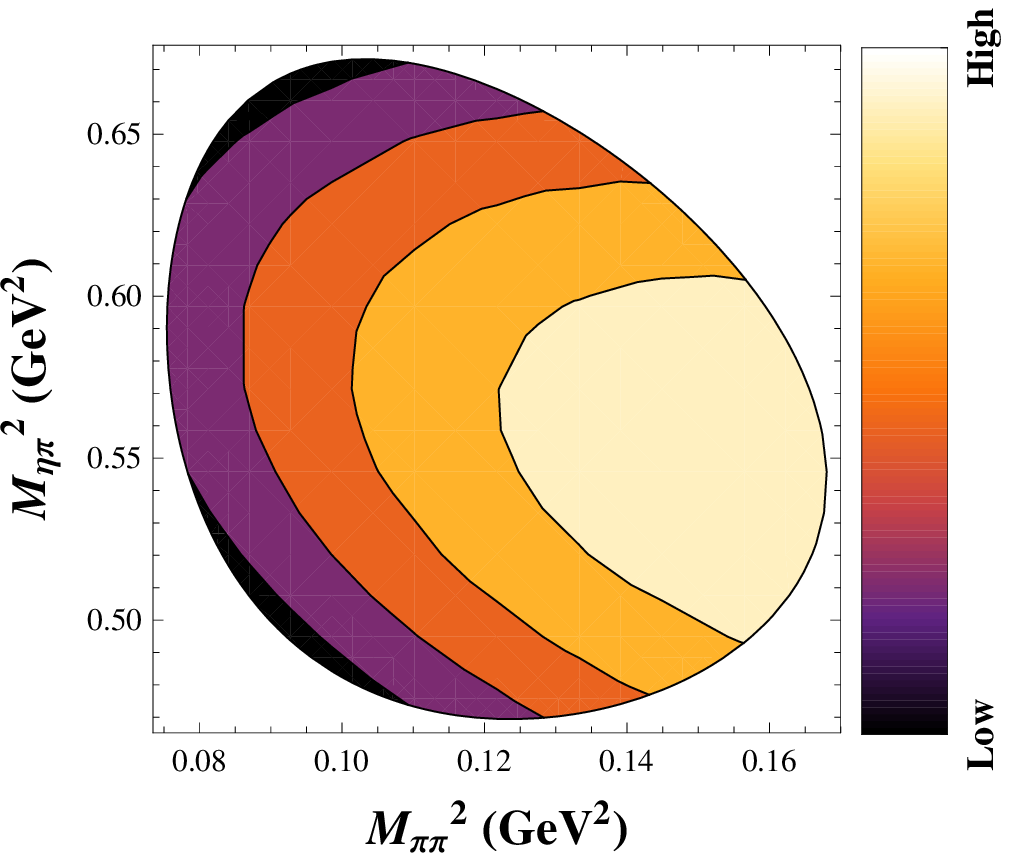}
\includegraphics[angle=0,clip,width=7cm]{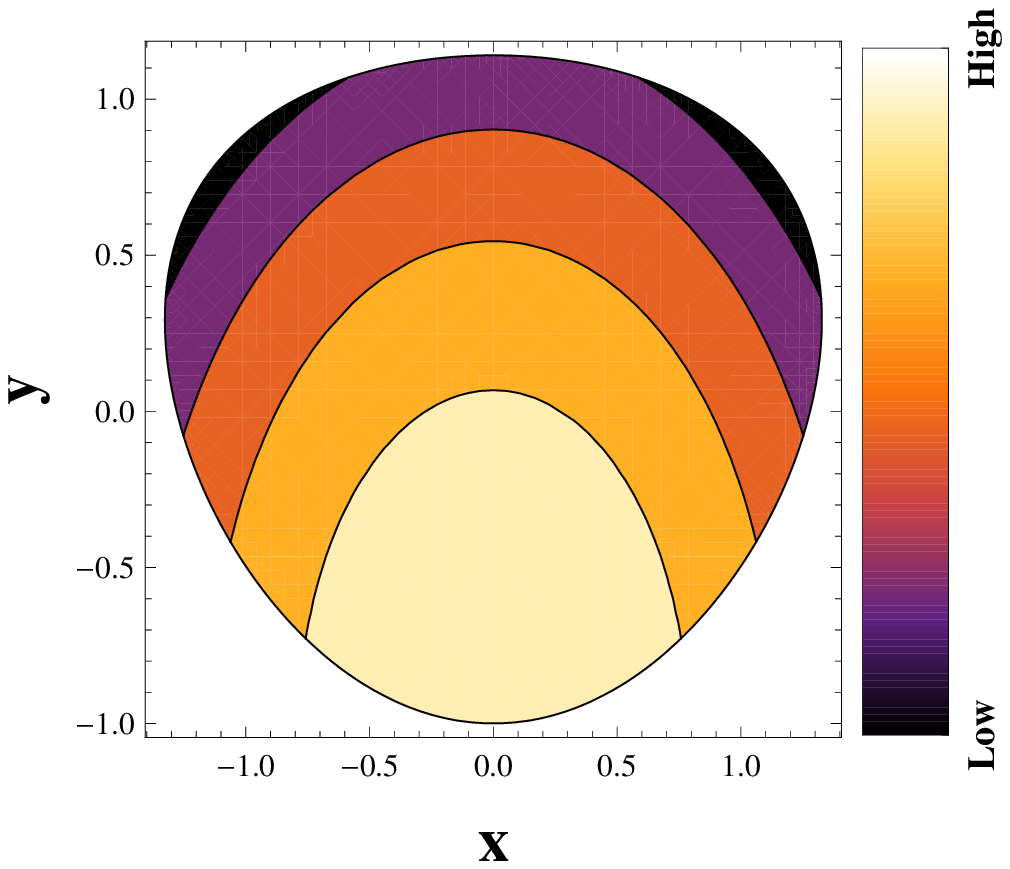}
\caption{
Dalitz plot distribution of $\eta^\prime\to\eta\pi\pi$ using Eq.~(\ref{amplLN})
supplemented by rescattering effects through Eq.~(\ref{amplB0unit}),
in terms of the invariant masses $M_{\pi\pi}^2$ and $M_{\eta\pi}^2$ (left)
and the kinematical variables $X$ and $Y$ (right).
Larger values are shown lighter.}
\label{fig.LNChPT-Dalitz}
}

\FIGURE{
\includegraphics[width=7cm]{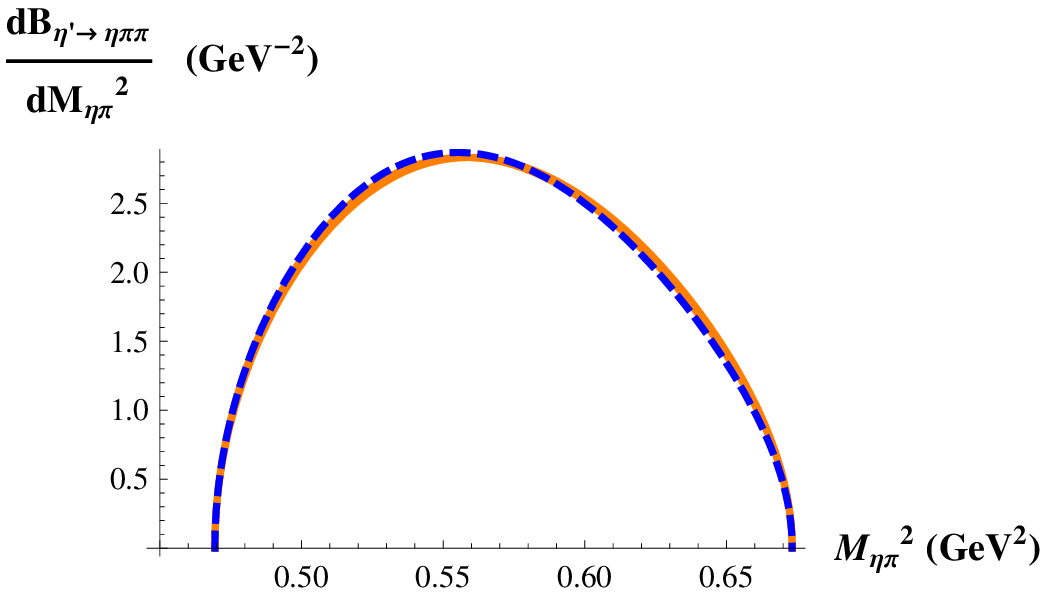}
\includegraphics[width=7cm]{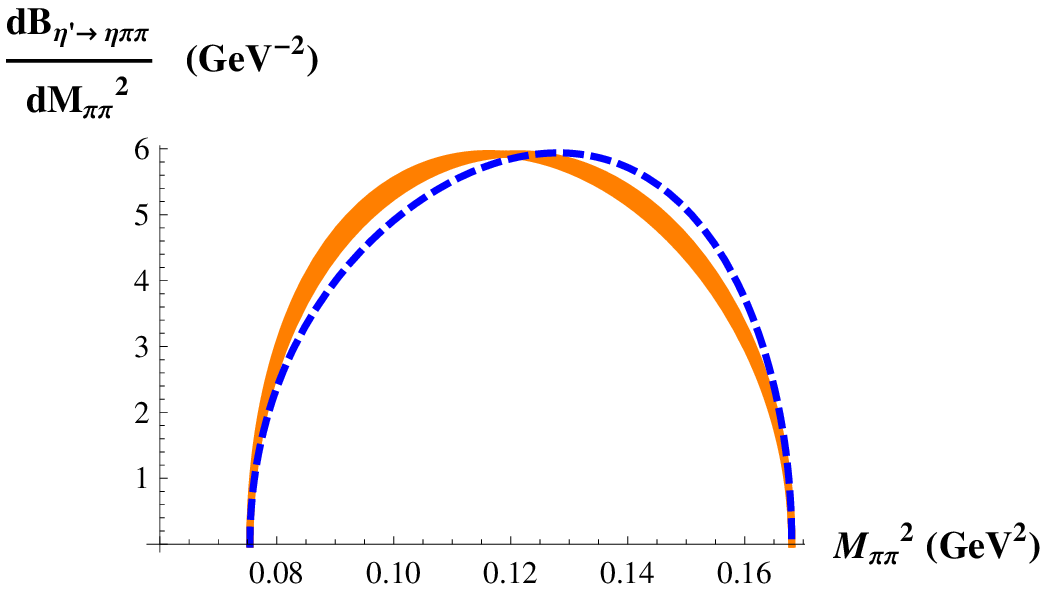}
\caption{
$M_{\eta\pi}^2$ (left) and $M_{\pi\pi}^2$ (right)
invariant mass spectra for the differential branching ratio.
The tree-level large-$N_C$ ChPT prediction from Eq.~(\ref{amplLN}) ---blue dashed line---
is compared to its unitarized counterpart via Eq.~(\ref{amplB0unit}) ---solid orange band.}
\label{Fig.InvMassLNChPT}
}

\section{Resonance Chiral Theory prediction}
\label{RChT}
As we discussed in the previous section, large-$N_C$ ChPT at NLO provides a successful prediction for the Dalitz parameters.
However, the comparison of Tables \ref{tab.dalitzparamLNcfit} and \ref{tab.Bunidalitzparam}   shows that final state interactions effects (NNLO in the large-$N_C$ ChPT counting)
are not negligible and must be properly incorporated.
One may wonder then whether other possible NNLO ChPT contributions could be relevant,
such as higher dimension local terms in the chiral Lagrangian
(\textit{e.g.}~${\cal O}(p^6)$ operators).
Previous studies have found that these local contributions can be saturated
by tree-level resonance exchanges~\cite{RChTb,RChTa,RChTc},
where the heavy meson propagators resum higher chiral orders.
Thus, we will make a thorough estimate of the impact of these higher chiral orders
by observing the contribution from intermediate resonant states.

\subsection{Framework and lowest order calculation}
Resonance Chiral Theory (RChT) is a description of the Goldstone and resonance interactions in a chiral invariant framework \cite{RChTb,RChTa,RChTc}.
The pseudo-Goldstones enter through the exponential realization
$u=\exp{\left(i\Phi/\sqrt{2}f\right)}$.
As the standard effective field theory momentum expansion is not valid in the presence of heavy resonance states, RChT takes the formal $1/N_C$ expansion as a guiding principle
\cite{largeNC}.
Thus, at large-$N_C$, the mesons get classified in $U(n_f)$ multiplets,
with $n_f=3$ the number of light-quark flavours.
The pseudo-Goldstone multiplet gathers then the octet of light pseudo-scalars
$(\pi,K,\eta_8)$ and the chiral singlet $\eta_1$, conforming a $U(3)$ nonet $\Phi$.
For convenience, the Lagrangian can be then organized
according to the number of resonance fields in the interaction terms,
\begin{equation}
\mathcal{L}_{\mathrm{R}\chi\mathrm{T}}=
\mathcal{L}^{\mathrm{GB}} + \mathcal{L}^{R_i} + \mathcal{L}^{R_iR_j} + \mathcal{L}^{R_iR_jR_k}
+\cdots\ ,
\label{rchtLagrangian}
\end{equation}
where $R_i$ stands for the resonance multiplets and the first term in the r.h.s.~of the equation
contains the operators without resonance fields,
\begin{equation}
 \mathcal{L}^{\mathrm{GB}}=\frac{f^2}{4}\langle u_\mu u^\mu +\chi_+\rangle\ ,
\end{equation}
with $u_\mu=i u^\dagger\partial_\mu U u^\dagger=u_\mu^\dagger$
(external sources are not included) and
$\chi_+=u^\dagger\chi u^\dagger + u\chi^\dagger u$.
For scalar resonances, $S(0^{++})$, the second term in Eq.~(\ref{rchtLagrangian})
is given by the operators~\cite{RChTa}
\begin{equation}
\mathcal{L}^S=c_d\langle S u_\mu u^\mu\rangle + c_m\langle S\chi_+\rangle\ ,
\label{1RLagrangian}
\end{equation}
where only the terms with a chiral tensor of order $p^2$ have been shown.
The resonance Lagrangian will be supplemented with kinetic and mass terms for the resonance fields.
The Lagrangian (\ref{rchtLagrangian}) also contains operators with vector meson fields,
but these do not contribute to $\eta^\prime\to\eta\pi\pi$ because of $G$-parity conservation.

A priori, one should also consider resonance operators in ${\cal L}_{\mathrm{R}\chi\mathrm{T}}$
with chiral tensors of arbitrary order\footnote{
The infinite tower of mesons contained in large-$N_C$ QCD
is often truncated to the lowest states in each channel with successful predictions for the
${\cal O}(p^4)$ and ${\cal O}(p^6)$ low-energy constants.
However, it has been pointed out in Ref.~\cite{PAs} that large discrepancies may occur
between the values of the masses and couplings of the full large-$N_C$ theory and
those from descriptions with a finite number of resonances.}.
For the case of scalar resonance operators,
one can prove through meson field redefinitions that the most general
scalar-pseudoscalar-pseudoscalar interaction in the chiral limit is provided by the
$c_d$ term in Eq.~(\ref{1RLagrangian}) \cite{cd+EoM}.
On the other hand, although nothing prevents the presence of ${\cal L}^{\rm GB}$
operators of order higher than $p^2$, within the large-$N_C$ limit,
a bounded behaviour of the $\eta^\prime\to\eta\pi\pi$ partial-wave amplitude at high energies requires the absence of ${\cal O}(p^4)$ operators and higher in ${\cal L}^{\rm GB}$.
Consequently, the sum ${\cal L}^{\rm GB}+\mathcal{L}^S$ contains all the pieces of the RChT Lagrangian that one would need to describe the $\eta^\prime\to\eta\pi\pi$ decay in the
large-$N_C$ limit.
Needless to say, the $U(1)_A$ anomaly contribution to the singlet mass must be also taken into account by means of the Lagrangian \cite{RChTc}
\begin{equation}
\label{eq.eta1-mass}
{\cal L}_{\eta_1}=\frac{f^2}{3}M_0^2\ln^2\left(\det{u}\right)=-\frac{1}{2}M_0^2\eta_1^2\ .
\end{equation}
This contribution,
which is not fixed by symmetry requirements alone and depends crucially on the dynamics of instantons,
is formally next-to-leading order in the large-$N_C$ expansion.
However, it is essential for the correct description of the $\eta^\prime$ and its decays.
Apart from this particularly large effect, any other NLO effect in $1/N_C$ is expected to be suppressed.

Concerning the $\eta^\prime\to\eta\pi\pi$ calculation in the framework of RChT,
we follow Sec.~\ref{NLOChPT} and express the amplitude in terms of
OZI-allowed and OZI-suppressed contributions.
At lowest order in large-$N_C$, the latter vanish,
${\cal M}_{\eta_s \eta_q \pi\pi}={\cal M}_{\eta_s \eta_s \pi\pi}=0$,
and ${\cal M}_{\eta^\prime\to\eta\pi\pi}=c_{qq}{\cal M}_{\eta_q \eta_q \pi\pi}$ is given by
\begin{eqnarray}
{\cal M}_{\eta^\prime\to\eta\pi\pi}^{\rm RChT}&=&
{c_{qq}\over f^2}\left[\frac{m_\pi^2}{2}
+\frac{1}{f_\pi^2}\frac{(c_d (s-m_{\eta^\prime}^2-m_\eta^2)+2c_m  m_\pi^2)
                                       (c_d (s-2 m_\pi^2)+2c_m m_\pi^2)}{M_{S}^2-s}
\right.\nonumber\\[1ex]
&&\left.
+\frac{1}{f_\pi^2}\frac{(c_d (t-m_{\eta^\prime}^2-m_\pi^2)+2 c_m m_\pi^2)
                                       (c_d (t-m_\eta^2-m_\pi^2)+2 c_m  m_\pi^2)}{M_{S}^2-t}
\right.\nonumber\\[1ex]
&&\left.
+\frac{1}{f_\pi^2}\frac{(c_d (u-m_{\eta^\prime}^2-m_\pi^2)+2 c_m m_\pi^2)
                                       (c_d (u-m_\eta^2-m_\pi^2)+2 c_m  m_\pi^2)}{M_{S}^2-u}\right]\ ,
\label{eq.RChT-amplitude}
\end{eqnarray}
where one has the contribution of the $I=1$ scalar resonance $a_0$ in the $t$- and $u$-channel
and the $I=0$ scalars $\sigma$ and $f_0$ in the $s$-channel.
$M_S$ is the mass of the scalar multiplet.
Although the Lagrangian (\ref{1RLagrangian}) does not include scalar mass splitting nor
$\sigma$-$f_0$ mixing,
these can be incorporated in the theory in a straightforward way through an operator of the form
$e^S_m\langle SS\chi_+\rangle$~\cite{Guo-aIJ,ms-split}.
Without loss of generality, one may consider different masses for the resonances in the scalar multiplet and a mixing scheme for the $\sigma$ and $f_0$.
In the heavy scalar mass limit, that is $p^2\sim m_P^2\ll M_S^2$,
the amplitude in Eq.~(\ref{eq.RChT-amplitude}) becomes
\begin{eqnarray}
{\cal M}_{\eta^\prime\to\eta\pi\pi}^{\rm RChT}&\longrightarrow&
{c_{qq}\over f^2}\left[\frac{m_\pi^2}{2}
+\frac{c_d^2}{f_\pi^2 M_S^2}(s^2+t^2+u^2-m_{\eta^\prime}^4-m_\eta^4-2m_\pi^4)
\right.\nonumber\\[1ex]
&&\left.
-\frac{2c_d c_m}{f_\pi^2 M_S^2}(m_{\eta^\prime}^2+m_\eta^2+2m_\pi^2)m_\pi^2
+\frac{12c_m^2}{f_\pi^2 M_S^2}m_\pi^4\right]\ ,
\end{eqnarray}
which turns out to be analogous to the large-$N_C$ ChPT amplitude in Eq.~(\ref{amplLN})
up to subleading contributions in $1/N_C$.
After using the low-energy constant relations
$3L_2+L_3=c_d^2/2M_S^2$, $L_5=c_d c_m/M_S^2$ and $L_8=c_m^2/ 2 M_S^2$ \cite{RChTa}, and the large-$N_C$ result $\Lambda_{1,2}=0$,
the two amplitudes are seen to be formally identical.

The largest contribution in Eq.~(\ref{eq.RChT-amplitude}) to the branching ratio
comes from the $c_d$ terms, which are proportional to the external momenta.
Everything else is proportional to $m_\pi^2$, being suppressed.
Nevertheless, even though the decay rate is essentially proportional to $c_d^4$,
the interference terms in the squared amplitude are not numerically negligible
and must be kept at the current level of experimental precision.
In addition to the branching ratio, the study of the shape of the differential decay width provides
important information.
For instance, the amplitude in Eq.~(\ref{eq.RChT-amplitude}) only depends on $X^2$
in the chiral limit,
\begin{equation}
\lim_{m_\pi\to 0}{\cal M}_{\eta^\prime\to\eta\pi\pi}^{\rm RChT}=
\frac{c_{qq}}{f^2}\frac{6c_d^2}{f_\pi^2}
\frac{\alpha_0 + \alpha_2 X^2}{\beta_0 + \beta_2 X^2}\ ,
\end{equation}
where $\alpha_n$ and $\beta_n$ are some polynomial combinations of just
$m_\eta$, $m_{\eta^\prime}$ and $M_S$.
This pure $X^2$ dependence was also found before in the large-$N_C$ ChPT framework
but with simpler polynomial combinations of pseudo-Goldstone masses.

For the numerical analysis,
we use the same input data we utilized in Sec.~\ref{NLOChPT} for the mixing parameters
and pseudoscalar masses, together with $M_S=980$ MeV for the scalar multiplet mass.
For the scalar couplings $c_d$ and $c_m$,
the latter, which is less relevant since it always appears through chiral suppressed contributions,
will be fixed via the high-energy scalar form factor constraint $4c_d c_m=f^2$
(indeed, $4c_d c_m\simeq f_\pi^2$ is taken) \cite{cdcm}.
Different values for these couplings are reported in the literature.
From the $a_0(980)$ decay width,
$c_d=26\pm 9$~MeV and $c_m=80\pm 21$~MeV~\cite{Guo-aIJ};
from low-energy constant resonance saturation,
$c_d=30\pm 10$~MeV and $c_m=43\pm 14$~MeV~\cite{RChTa,cdcm2};
from the $I=1/2$ $K\pi$ $s$-wave scattering amplitude analysis,
$c_d=25$~MeV and $c_m=77$~MeV~\cite{cdcm2}, and
from the $I=3/2$ $K\pi$ $s$-wave,
$c_d=13$~MeV and $c_m=85$~MeV~\cite{cdcm2}.
Notice that all these results obey rather well the theoretical constraint $4 c_d c_m=f^2$.

We mentioned that the largest contribution to the branching ratio comes from the $c_d$ terms.
Indeed, it is to a large extent proportional to the fourth power of $c_d$.
Therefore, owing to the large uncertainty in $c_d$, we prefer to fix the value of this coupling from the experimental branching ratio and predict the Dalitz plot parameters.
In doing so, one gets $c_d\simeq 28$~MeV, in agreement with most of the reported values.
A first prediction of the Dalitz plot parameters is shown in the second column of
Table \ref{tab.LO-RChT}.
While $a$ and $d$ are in nice agreement with the VES results, $b$ is very far.
We also calculate certain combinations of these parameters that were found to be independent of the chiral couplings in the large-$N_C$ ChPT framework.
In the case of RChT, even though the cancelation of couplings is not complete,
there is not a sizable dependence on the value of $c_d$.
In particular, one finds $a/d=2.1$ and $\kappa_{21}/2\kappa_{40}=2.4$,
to be compared with $a/d=\kappa_{21}/2\kappa_{40}=3.4$ in Eq.~(\ref{dalitzparamrelations}).
The RChT result for the first ratio agrees better with the measurement from VES,
$a/d|_{\rm exp}=1.55\pm 0.42$, than the large-$N_C$ ChPT prediction.

\TABLE{
\begin{tabular}{|l|c|c|c|c|}
  \hline
   Parameter & $J=0$	& $J=0,2$	&
   GAMS-4$\pi$	 \cite{GAMSexp}	& VES  \cite{VESexp}\\
   \hline
   $a[Y]$					& $-0.119$	& $-0.160$	& $-0.066\pm 0.016$
   									& $-0.127\pm 0.018$\\
   $b[Y^2]$				& $+0.001$   	& $-0.001$	& $-0.063\pm 0.028$
   									& $-0.106\pm 0.031$\\
   $d[X^2]$				& $-0.056$	& $-0.070$	& $+0.018\pm 0.078$
   									& $-0.082\pm 0.019$\\
   $\kappa_{21}[X^2Y]$		& $-0.006$	& $-0.003$	& ---		& ---		\\
   $\kappa_{40}[X^4]$		& $-0.001$	& $-0.001$	& ---		& ---		\\
   \hline
   $c_d$ (MeV)				& 28			& 26			&		&		\\
   \hline
\end{tabular}
\caption{Dalitz plot parameters obtained from the RChT amplitude in Eq.~(\ref{eq.RChT-amplitude})
(second column) and with the addition of $J=2$ resonances via Eq.~(\ref{amplLNT})
(third column).}
\label{tab.LO-RChT}
}

In addition to the contribution of scalar mesons,
we also consider the impact of $J=2$ resonances.
In the previous large-$N_C$ ChPT analysis, the $J=2$ partial wave was shown to be crucial to properly recover the Dalitz plot parameters.
Still, as the mass of the lightest tensor multiplet is roughly $M_{f_2}=1.2$ GeV,
one may just consider its leading effect in $1/M_{f_2}^2$ rather than the whole non-local resonance propagator structure.
Thus, it induces a contribution that is identical to the $3 L_2+L_3$ term in Eq.~(\ref{amplLN}),
\begin{equation}
\label{amplLNT}
{\cal M}_{\eta^\prime\to\eta\pi\pi}^T=
{c_{qq}\over f^2}
 \frac{2(3L_2^T+ L_3^T)}{f_\pi^2}(s^2+t^2+u^2-m^4_{\eta^\prime}-m^4_{\eta}-2 m^4_{\pi})\ .
 \end{equation}
The $J=2$ resonance contributions to the ${\cal O}(p^4)$ LECs,
$3 L_2^T+L_3^T= g_{f_2}^2/3 M_{f_2}^2=0.16\cdot 10^{-3}$
were estimated in Ref.~\cite{Ecker-tensor},
after imposing high-energy constraints on $\pi\pi$-scattering.
When the tensor contribution is taken into account, the value of the scalar coupling obtained from the experimental branching ratio is $c_d\simeq 26$~MeV.
The predicted Dalitz plot parameters in this case are shown in the third column of
Table \ref{tab.LO-RChT}.
The values for $a$ and $d$ are still in reasonable agreement with VES, but $b$ continues to be very far.

\subsection{Resonance width effects}

When comparing Tables~\ref{tab.dalitzparamLNcfit} and~\ref{tab.LO-RChT},
one can see that RChT at  tree-level improves considerably the prediction for the $a$ parameter from NLO large-$N_C$ ChPT.
Nonetheless, in both cases, one finds a far too small value for $b$.
This issue was solved in large-$N_C$ ChPT by considering the FSI impact from meson loops through a convenient unitarization scheme.
It is then desirable to incorporate loops in the resonance description in order to refine our prediction, the same we did in the ChPT framework.
The first obvious FSI effect is that the scalar mesons gain a non-zero width.
In particular, the scalar isoscalar $\sigma$ meson becomes broad.
Although a detailed study of these NLO effects in $1/N_C$ is beyond the scope of this work,
we perform some estimates of the theoretical uncertainty associated with this large-$N_C$ description.

A first estimate is provided by the inclusion of a $\sigma$-$f_0(980)$ splitting
and a self-energy in the $s$-channel propagator of the $\sigma$,
by means of the substitution in Eq.~(\ref{eq.RChT-amplitude}),
\begin{equation}
\label{eq.S-propagator}
\frac{1}{M_S^2-s}\longrightarrow
\frac{\sin^2\phi_S}{M_{f_0}^2-s}+
\frac{\cos^2\phi_S}{M_\sigma^2-s-c_\sigma s^k \bar{B}_0(s,m_\pi^2,m_\pi^2)}\ ,
\end{equation}
with $M_{f_0}=980$~MeV, $\phi_S=-8^\circ$~\cite{Escribano:2006mb} and the parameters
$M_\sigma$ and $c_\sigma$ tuned such that one recovers the right position for the $\sigma$ pole,  $M_\sigma^{\rm pole}= 441^{+16}_{-8}$~MeV and
$\Gamma_\sigma^{\rm pole}=544^{+18}_{-25}$~MeV~\cite{CCL-sigma}.
The function $\bar{B}_0$ is the subtracted two-point Feynman integral written in
Eq.~(\ref{eq.B0-def}) with $C=2$ and $\rho(s)=\sqrt{1-4 m_\pi^2/s}$.
The power behaviour $k=0$ produces an unphysical bound state in the first Riemann sheet very close to the $\pi\pi$ threshold from below,
which enhances unnaturally the amplitude and leads to $c_d\simeq 10$~MeV.
This case seems to be disfavoured from the phenomenological point of view and is discarded.
For $k=1$, the amplitude reproduces the $\sigma$ pole and fixes $c_d\simeq 26$~MeV.
Power behaviours equal or higher than $k=2$ are unable to generate the $\sigma$ pole at the right position, and are also discarded.
The predicted Dalitz plot parameters for $k=1$ are shown in Table~\ref{tab.width+RChT}.
As seen, the value of $b$ moves into the right direction, $a$ increases with respect to its value without including $\sigma$-width effects, and the rest of parameters present small variations
(see Table \ref{tab.LO-RChT} for comparison).
One can take also into account the $f_0(980)$ and $a_0(980)$ widths in a similar way\footnote{
For the $f_0$ and $a_0$ scalar mesons,
the important piece of the self-energy is its imaginary part,
being the real part of its corresponding logarithm almost negligible in comparison with the dominant contribution $M_S^2- k^2$ (with $k^2=s, t, u$, depending on the channel).
In any case, we consider the full self-energy function as done in Eq.~(\ref{eq.S-propagator}).
The function $\bar{B}_0(k^2,m_\pi^2,m_\eta^2)$ appearing in the $t$- and $u$-channel
can be found in Ref.~\cite{GL85}},
but their effects are seen to be clearly subdominant as compared to the impact of a broad $\sigma$
and they slightly change the Dalitz plot parameters obtained before.

\TABLE{
\begin{tabular}{|l|c|c|c|c|}
  \hline
   Parameter 				& $J=0$		& $J=0$		&
   GAMS-4$\pi$	 \cite{GAMSexp}	& VES  \cite{VESexp}\\
   & $+\sigma$ width	& $+(\sigma,f_0,a_0)$ widths	&	& \\
   \hline
   $a[Y]$					& $-0.161$	& $-0.168$	& $-0.066\pm 0.016$
   									& $-0.127\pm 0.018$\\
   $b[Y^2]$				& $-0.034$   	& $-0.035$	& $-0.063\pm 0.028$
   									& $-0.106\pm 0.031$\\
   $d[X^2]$				& $-0.055$	& $-0.056$	& $+0.018\pm 0.078$
   									& $-0.082\pm 0.019$\\
   $\kappa_{21}[X^2Y]$		& $-0.005$	& $-0.006$	& ---		& ---		\\
   $\kappa_{40}[X^4]$		& $-0.001$	& $-0.001$	& ---		& ---		\\
   \hline
   $c_d$ (MeV)				& 28			& 26			&		&		\\
   \hline
\end{tabular}
\caption{Dalitz plot parameters obtained from the RChT amplitude in Eq.~(\ref{eq.RChT-amplitude})
supplemented by $\sigma$-width effects (second column) and with the addition of $f_0$- and $a_0$-width effects (third column).}
\label{tab.width+RChT}
}

The estimate of the FSI effects provided here is obviously model dependent,
as we have introduced \textit{ad hoc} a splitting and a self-energy for the scalar multiplet.
The splitting can be easily introduced through the corresponding terms in the Lagrangian,
an issue studied in detail in Ref.~\cite{ms-split}.
On the other hand, the resummation of the one-loop self-energy effects is also justified,
even for the case of the broad $\sigma$.
Higher order effects as multimeson channels are negligible below 1 GeV and the one-loop amplitude seems to provide the relevant information in the physical range under study.
In any case, a self-energy must be appropriately resummed in the neighbourhood of the
resonance pole \cite{RChT-width,RGE}.
Nevertheless, as we are aware of the level of model dependence of the present procedure,
in the next subsection we use the $N/D$ unitarization method to incorporate meson rescattering effects in a less dependent way and compare both approaches.

\subsection{$\pi\pi$ final state interactions}

As in large-$N_C$ ChPT, here we consider FSI by demanding unitarity in the $\pi\pi$ channel.
We have already seen that this is by far the most important rescattering effect due to the little impact of the $a_0$ width (see Table~\ref{tab.width+RChT}).
We use again the $N/D$ method to unitarize the tree level RChT amplitude
in Eq.~(\ref{eq.RChT-amplitude}) trough the expression in Eq.~(\ref{amplB0unit}).
In this manner, provided the absorptive cuts in the $t$- and $u$-channel are neglected,
only the $\pi\pi$-scattering amplitude in the $s$-channel has to be specified
and not the precise way the resonances get their width.
The scalar coupling $c_d$ and the constant $C$ appearing in the $B_0$ integral
of Eq.~(\ref{eq.B0-def}) will be fixed from the experimental branching ratio and the $a$ parameter.
The predictions for the other Dalitz plot parameters are given in
Table~\ref{tab.unitariza-RChT}.
We consider these results as the most reliable RChT estimate.
The comparison of the third column in Table~\ref{tab.width+RChT}
and the second column in Table~\ref{tab.unitariza-RChT}
shows a fair agreement between the two FSI approaches,
resonance propagator modification and $N/D$ unitarization.
The largest difference is in the $a[Y]$ parameter, which is taken as an input in the $N/D$ method.

\TABLE{
\begin{tabular}{|l|c|c|c|c|}
  \hline
   Parameter 				& $J=0$		& $J=0,2$		&
   GAMS-4$\pi$	 \cite{GAMSexp}	& VES  \cite{VESexp}\\
   & +$N/D$ unitar.	& +$N/D$ unitar.	&	& \\
   \hline
   $b[Y^2]$				& $-0.030(1)$   	& $-0.033(1)$	& $-0.063\pm 0.028$
   									& $-0.106\pm 0.031$\\
   $d[X^2]$				& $-0.057(1)$	& $-0.072(1)$	& $+0.018\pm 0.078$
   									& $-0.082\pm 0.019$\\
   $\kappa_{21}[X^2Y]$		& $-0.010(1)$	& $-0.009(2)$	& ---		& ---		\\
   $\kappa_{40}[X^4]$		& $+0.001(1)$	& $+0.001(1)$	& ---		& ---		\\
   \hline
   $c_d$ (MeV)				& 27(1)		& 25(1)		&		&		\\
   $C$-constant 			& $2.1 \leq C \leq 3.4$
   						& $1.4 \leq C \leq 2.7$		&		&		\\
   \hline
\end{tabular}
\caption{Dalitz plot parameters obtained from the RChT amplitude in Eq.~(\ref{eq.RChT-amplitude})
supplemented by rescattering effects via Eq.~(\ref{amplB0unit}) (second column) and with the addition of $J=2$ resonances (third column).}
\label{tab.unitariza-RChT}
}

To summarize, if one remains at tree level there is no way to increase the prediction for the
$b[Y^2]$ parameter from a value very close to zero to the size of measured values.
The same was found in large-$N_C$ ChPT.
This situation is greatly improved as soon as one incorporates the rescattering effects,
see Table~\ref{tab.unitariza-RChT}.
Now, the $b$ parameter is compatible with the GAMS measurement
and not far from the VES result.
On the other hand, the $d[X^2]$ parameter seems to be rather stable,
both in large-$N_C$ ChPT and RChT,
either at tree level or including rescattering effects.
It remains compatible with the various experimental values,
in particular, with the most precise measurement reported by VES.

Finally,
we display in Fig.~\ref{fig.RChT-Dalitz},
the Dalitz plot distribution of $\eta^\prime\to\eta\pi\pi$ using the RChT amplitude
in Eq.~(\ref{eq.RChT-amplitude}) supplemented by rescattering effects through
Eq.~(\ref{amplB0unit}),
and in Fig.~\ref{fig.marginal-RChT},
the corresponding invariant mass spectra from
Eq.~(\ref{eq.RChT-amplitude}) ---blue dashed line--- and Eq.~(\ref{amplB0unit})
---solid orange band.
In Fig.~\ref{fig.marginal-RChT},
we use as input the values shown the third columns of
Tables \ref{tab.LO-RChT} and \ref{tab.unitariza-RChT}
for the lowest order calculation and its $N/D$ unitarized counterpart, respectively.

\FIGURE{
\includegraphics[angle=0,clip,width=7cm]{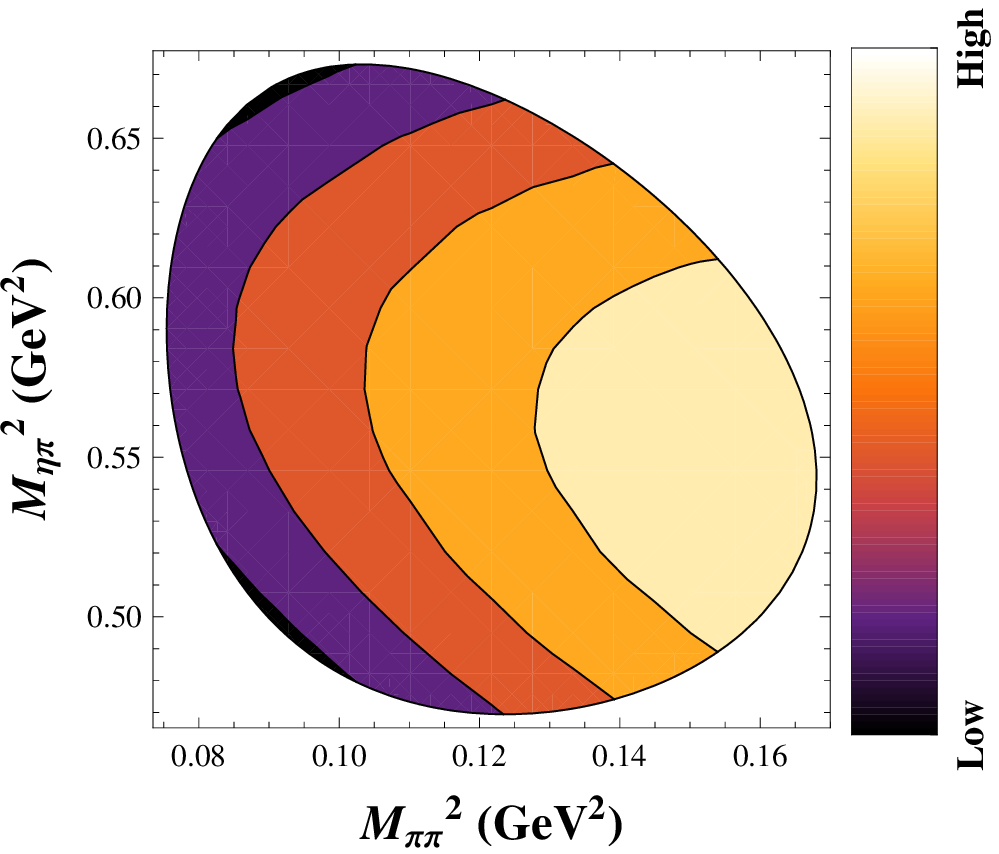}
\includegraphics[angle=0,clip,width=7cm]{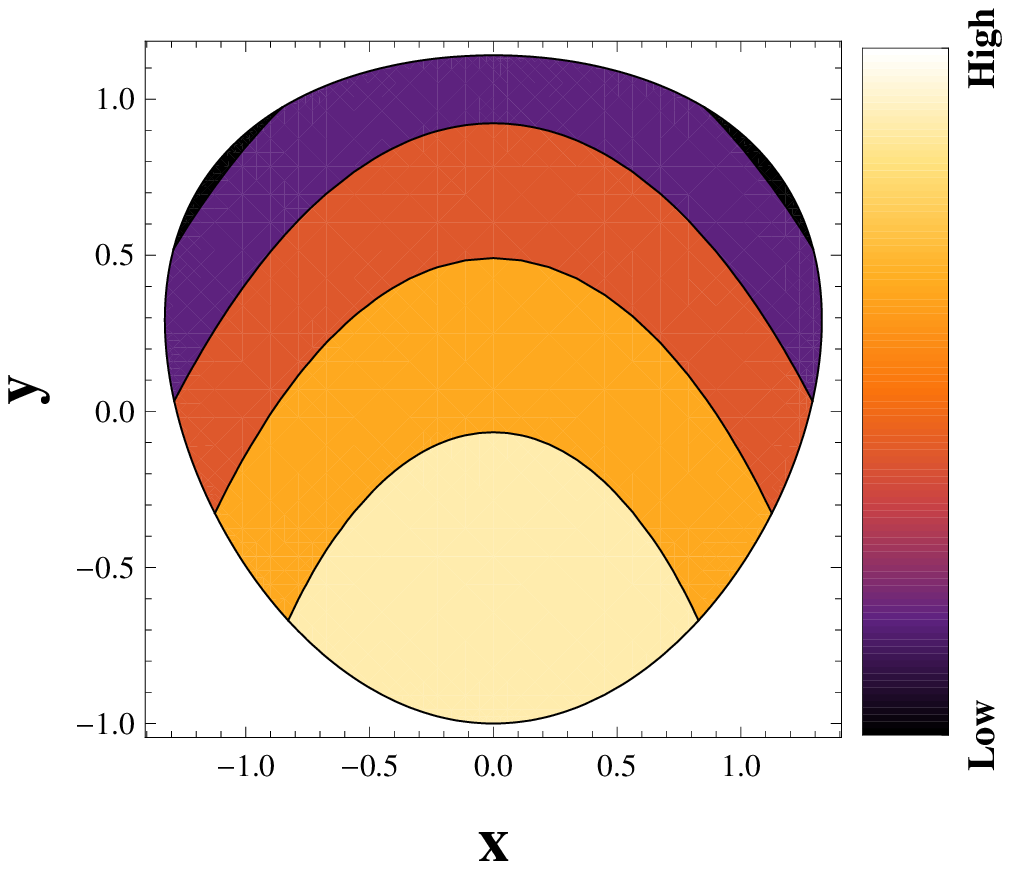}
\caption{
Dalitz plot distribution of $\eta^\prime\to\eta\pi\pi$ using Eq.~(\ref{eq.RChT-amplitude})
supplemented by rescattering effects through Eq.~(\ref{amplB0unit}),
in terms of the invariant masses $M_{\pi\pi}^2$ and $M_{\eta\pi}^2$ (left)
and the kinematical variables $X$ and $Y$ (right).}
\label{fig.RChT-Dalitz}
}

\FIGURE{
\includegraphics[width=7cm]{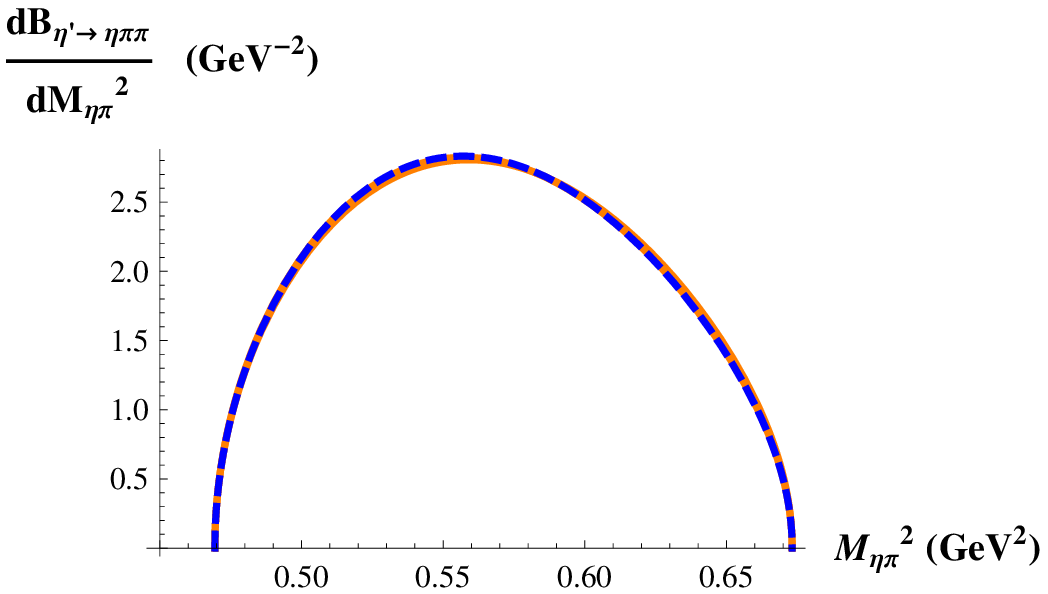}
\includegraphics[width=7cm]{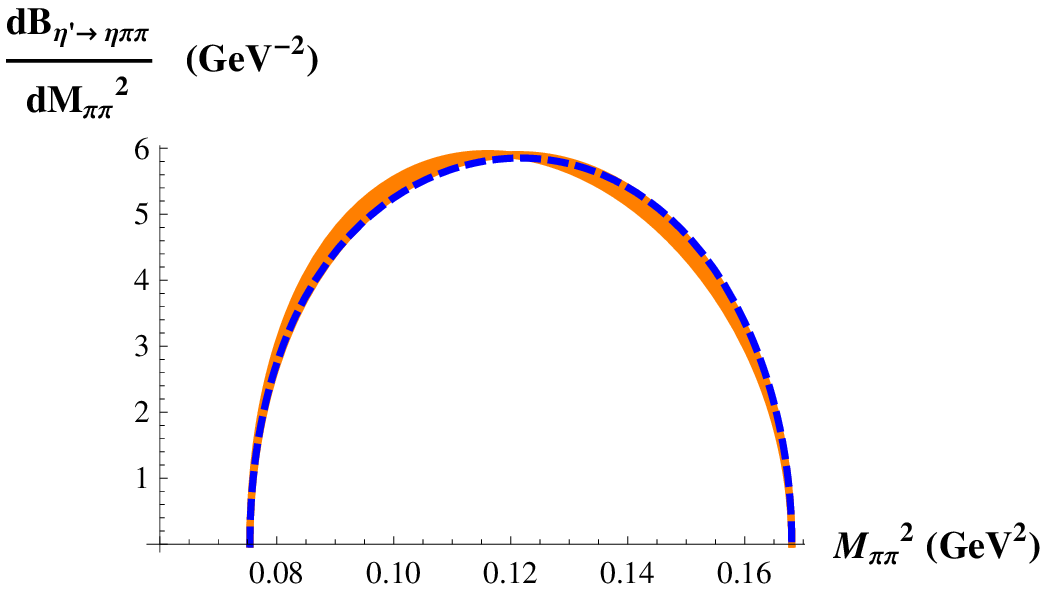}
\caption{
$M_{\eta\pi}^2$ (left) and $M_{\pi\pi}^2$ (right)
invariant mass spectra for the differential branching ratio.
The tree-level RChT prediction from Eq.~(\ref{eq.RChT-amplitude}) ---blue dashed line---
is compared to its unitarized counterpart via Eq.~(\ref{amplB0unit}) ---solid orange band.}
\label{fig.marginal-RChT}
}

\section{Discussion and conclusions}
\label{conclusions}
In this work, we have studied the hadronic decays $\eta^\prime\to\eta\pi\pi$
in the frameworks of large-$N_C$ Chiral Perturbation Theory,
at lowest and next-to-leading orders, and
Resonance Chiral Theory in the leading $1/N_C$ approximation.
In both cases, we have also considered higher order effects such as $\pi\pi$
final state interactions through detailed unitarization procedures.
The inclusion of finite-width effects in the case of RChT is also discussed.
In addition, and for the benefit of present and future experimental analyses,
we have computed the Dalitz plot distribution and the differential branching ratio.

In large-$N_C$ ChPT, the next-to-leading order calculation in Eq.~(\ref{amplLN})
represents a clear improvement with respect to the current algebra expression
in Eq.~(\ref{amplLO}), leading order in large-$N_C$ ChPT,
and yields a prediction for the branching ratio of the right size.
The RChT computation in Eq.~(\ref{eq.RChT-amplitude}) also constitutes an improvement
over the current algebra result.
However, the two approaches fail to reproduce the Dalitz plot parameters $a[Y]$ and $b[Y^2]$,
both associated with $s$-channel exchanges and then related to isoscalar scalar resonances.
This points out the importance of $\sigma$ meson effects in this channel.
On the contrary, the contributions from the $t$- and $u$-channel are less problematic.
The predictions for the parameter $d[X^2]$ in the different frameworks are relatively consistent among themselves and agree with the experimental measurements.

\TABLE{
{\small
\begin{tabular}{|l|c|c|c|c|}
  \hline
   Parameter 					& Large-$N_C$ ChPT		& RChT($J=0,2$)
   & GAMS-4$\pi$ \cite{GAMSexp}	& VES  \cite{VESexp}\\
   & +$N/D$ unitar.				& +$N/D$ unitar.			&		& \\
   \hline
   $a[Y]$					& $-0.098(48)^\dagger$	& $-0.098(48)^\dagger$
   						& $-0.066\pm 0.016$	& $-0.127\pm 0.018$\\
   $b[Y^2]$				& $-0.050(1)$   	& $-0.033(1)$	& $-0.063\pm 0.028$
   									& $-0.106\pm 0.031$\\
   $d[X^2]$				& $-0.092(8)$	& $-0.072(1)$	& $+0.018\pm 0.078$
   									& $-0.082\pm 0.019$\\
   $\kappa_{21}[X^2Y]$		& $+0.003(2)$	& $-0.009(2)$	& ---		& ---		\\
   $\kappa_{40}[X^4]$		& $+0.002(1)$	& $+0.001(1)$	& ---		& ---		\\
   \hline
   $(3L_2+L_3)\cdot 10^3$	& 1.0(1)		& ---			&		&		\\
   $c_d$ (MeV)				& ---			& 25(1)		&		&		\\
   $C$-constant 			& $-1.4 \leq C \leq 0.9$
   						& $1.4 \leq C \leq 2.7$		&		&		\\
   \hline
\end{tabular}
\caption{Final results for the Dalitz plot parameters obtained in the frameworks of
large-$N_C$ ChPT (second column) and
RChT with the addition of $J=2$ resonances (third column), respectively,
after the unitarization of the corresponding amplitudes through the $N/D$ method.
The $\eta^\prime\to\eta\pi\pi$ experimental branching ratio and the Dalitz plot parameter $a$
have been used as inputs$^\dagger$ to fix the relevant couplings and the constant of the unitarization formula.}
\label{tab.final-results}
}}

Our results can be further improved through an appropriate unitarization of the amplitudes
in order to account for the final state interactions in the $\pi\pi$ system.
The best estimates are achieved with the $N/D$ unitarization method.
The final results for the Dalitz plot parameters using this method are summarized in
Table~\ref{tab.final-results}.
We show the results for \textit{set3} of low-energy constants in the case of large-$N_C$ ChPT
and including the effects of scalar and tensor resonances for the case of RChT.
For the sake of comparison, we also present the measured values.
All Dalitz plot parameters lie now within the experimental range.
The differences between the predictions of the two frameworks in Table~\ref{tab.final-results}
should be taken as the systematic uncertainty of our analysis.
In any case, it is much smaller than the dispersion exhibited by the measurements.
Preliminary results from BES-III seem to provide much more precise determinations,
$a=-0.047\pm 0.011\pm 0.003$, $b=-0.069\pm 0.019\pm 0.009$ and
$d=-0.073\pm 0.012\pm 0.003$~\cite{preliminary-BESIII}.
If confirmed, this would favour our predictions with respect to other theoretical studies
which predict, for instance, a very small and positive $d$ parameter
(see Table~\ref{tablepar} and Ref.~\cite{Borasoy:2005du}).

In conclusion,
the large-$N_C$ ChPT framework seems to be suitable for describing
$\eta^\prime\to\eta\pi\pi$ decays even though higher next-to-next-to-leading chiral orders
turn out to be relevant.
The RChT approach tries to cure this problem since part of the higher order local terms in ChPT
are now resummed through the exchange of heavy resonances.
We have seen that the $a_0(980)$ contribution is dominant but, at the same time,
the $\sigma$ contribution is also essential to agree with experiment.
However, we find that heavier states, which are not present in the original RChT framework,
are not negligible.
In particular, the lowest lying tensor resonances produce a noticeable effect in the RChT estimates.
Therefore, although RChT resums higher chiral orders, at this level of precision, it requires a much more detailed knowledge of the resonance content and properties.
On the other hand, in the large-$N_C$ ChPT approach,
the effect of any possible heavy state is encoded in the low-energy constants and hence
the resonance spectrum does not need to be known in detail.
For these reasons, we consider both frameworks, large-$N_C$ ChPT and RChT,
to be complementary,
where each one has its own advantages and disadvantages.
We hope future experiments to be able to distinguish the most convenient framework for the
study of this and other $\eta^\prime$ decays.

\acknowledgments
This work was supported in part by the Juan de la Cierva program (J.J.S.-C.),
the Ministerio de Ciencia e Innovaci\'on under grants
CICYT-FEDER-FPA2008-01430, FPA2008-04158-E, FPA2006-05294 and ACI2009-1055,
the EU Contract No.~MRTN-CT-2006-035482  ``FLAVIAnet'',
the European Commission under the 7th Framework Programme through the
'Research Infrastructures' action of the 'Capacities' Programme Call:
FP7-INFRA\-STRUCTURES-2008-1 (Grant Agreement N. 227431),
the Spanish Consolider-Ingenio 2010 Programme CPAN (CSD2007-00042),
the Generalitat de Catalunya under grant SGR2009-00894
and the Junta de Andaluc\'{\i}a (Grants P07-FQM 03048 and P08-FQM 101).
P.M. and J.J.S.-C. thank IFAE at the Universitat Aut\`{o}noma de Barcelona for the hospitality.
R.E. thanks S.~Peris for a careful reading of the manuscript.
We also thank  G.~Colangelo, B.~Kubis and J.R.~Pel\'aez for useful comments and suggestions,
and C.P.~Shen from the BESIII collaboration for calling our attention to their recent results on
$\eta^\prime\to\eta\pi^+\pi^-$ \cite{preliminary-BESIII}.
We would like to dedicate this work to the memory of Joaquim Prades.

\appendix

\section{Partial waves and unitarity}
\subsection{Partial-wave projection}
\label{pwp}
The projection of a given $\eta^\prime\to\eta\pi\pi$ invariant amplitude, ${\cal M}(s,t,u)$,
into partial waves is
\begin{equation}
{\cal M}(s,t,u)=\sum_J 32\pi (2J+1) P_J(\cos\theta_\pi){\cal M}_J(s)\ ,
\label{eq.PWdecomp}
\end{equation}
where $P_J$ is the $J$th Legendre polynomial, $\theta_\pi$ is the angle of $\mathbf{p}_{\pi^+}$ with respect to $\mathbf{p}_{\eta}$ in the $\pi\pi$ rest frame, and
\begin{equation}
\label{partialwave}
{\cal M}_J(s)={1\over 32\pi}
\frac{s}{\lambda(s,m_\eta^2,m_{\eta^\prime}^2)^{1/2}\lambda(s,m_\pi^2,m_\pi^2)^{1/2}}
\int_{t_{\rm min}}^{t_{\rm max}}dt P_J(\cos\theta_\pi){\cal M}(s,t,u)\ ,
\end{equation}
with
\begin{eqnarray}
\cos\theta_\pi &=&
-\frac{s (m_{\eta^\prime}^2+m_\eta^2+2 m_\pi^2-s-2t)}
{\lambda(s,m_\eta^2,m_{\eta^\prime}^2)^{1/2}\lambda(s,m_\pi^2,m_\pi^2)^{1/2}}\ ,\\[1ex]
t_{\rm max(min)} &=&
\frac{1}{2}
\left[m_{\eta^\prime}^2+m_\eta^2+2 m_\pi^2-s\pm
\frac{\lambda(s,m_\eta^2,m_{\eta^\prime}^2)^{1/2}\lambda(s,m_\pi^2,m_\pi^2)^{1/2}}{s}\right]\ ,
\end{eqnarray}
and $\lambda(x,y,z)=x^2+y^2+z^2-2x y-2x z-2y z$.
For a $\pi\pi$-scattering amplitude of isospin $I$,
its decomposition into partial waves is
\begin{equation}
\label{amplunitB0}
T^I(s,t,u)=\sum_J 32\pi (2J+1) P_J(1+\frac{2t}{s-4m_{\pi}^2}){\cal T}^I_J(s)\ ,
\end{equation}
with
\begin{equation}
{\cal T}^I_J(s)=\frac{1}{32\pi}\frac{1}{s-4 m_\pi^2}
\int^0_{4m_{\pi}^2-s}dt\, P_J(1+\frac{2t}{s-4m_{\pi}^2})T^I(s,t,u)\  .
\end{equation}
For the $I=0$ amplitude
$T^0(s,t,u)=3A(s,t,u)+A(t,s,u)+ A(u,t,s)$, with $A(s,t,u)=T_{\pi^+\pi^-\to\pi^0\pi^0}$,
the $J=0,2$ partial waves in large-$N_C$ ChPT are found to be
\begin{eqnarray}
\label{eq.Tpi00}
{\cal T}^0_0(s) &=&
\frac{s-m_{\pi}^2/2}{16\pi f_\pi^2}+\frac{1}{24\pi f_\pi^4}
\left[(55 L_2+17 L_3)s^2-8(40 L_2+11 L_3) s m_{\pi}^2
\right.\nonumber\\[1ex]
&&\left.
+10 (58 L_2+14 L_3 - 3L_5 + 6 L_8) m_{\pi}^4\right]\ ,\\[1ex]
{\cal T}^0_2(s) &=&
\frac{1}{120\pi f_\pi^4}(5L_2+L_3)(s-4m_{\pi}^2)^2\ .
\label{eq.Tpi02}
\end{eqnarray}

\subsection{Unitarization of final state interactions}
\label{unitarization}
Similar to meson-meson scattering amplitudes,
it is possible to use the optical theorem to write down a series of unitarity relations for production amplitudes.
Nonetheless, contrary to scattering, where in the physical region there are only absorptive cuts in the $s$-channel, in production processes one may also have unitarity cuts on the $t$- and $u$-channel within the physical phase-space.
Thus, strictly speaking, these channels are present in the unitarity relation.
However, for the reasons explained in the main text, only $s$-channel unitarity is considered here.
The unitarity relation taking into account $\pi\pi$ final state interactions in the $s$-channel is written as
\begin{equation}
\label{eq.unitarity}
\mbox{Im}{\cal M}_J(s)=\rho(s){\cal T}^0_J(s)^*{\cal M}_J(s)\ ,
\end{equation}
with
$\rho(s)=\lambda(s,m_\pi^2,m_\pi^2)^{1/2}/s=\sqrt{1-4m_\pi^2/s}$.
In Eq.~(\ref{eq.unitarity}), it is made explicit that the two-pion system in the final state of
$\eta^\prime\to\eta\pi\pi$ is always $I=0$ (assuming isospin conservation).
There are several ways to satisfy the former unitary relation.
The simplest one is based on the $K$-matrix  formalism where
\begin{equation}
\label{amplkmatrix}
{\cal M}_J(s)|_{\rm K-matrix}=
\frac{{\cal M}_J(s)|_{\rm tree}}{1-i\rho(s){\cal T}^0_J(s)|_{\rm tree}}\ ,
\end{equation}
with ${\cal M}_J(s)|_{\rm tree}$ and ${\cal T}^0_J(s)|_{\rm tree}$ calculated at tree level and thus real.
Through this formalism one incorporates the imaginary part of the pion loops.
However, it misses the real part of the logarithms that appear, for instance, in the ChPT calculation at the loop level.
One may think of completing the imaginary part $i\rho(s)$
appearing in the $K$-matrix  formalism with the full logarithm that shows up in the two-propagator Feynman integral
\begin{equation}
\label{eq.B0-def}
16\pi^2 B_0(s)=C-\rho(s)\log\frac{\rho(s)+1}{\rho(s)-1}\ ,
\end{equation}
with Im$B_0(s)=\rho(s)/16\pi$ for $s>4m_\pi^2$.
This can be then incorporated in a solution of the unitarity relation of a form similar to the $N/D$ method~\cite{Oset-ND,Palomar-ND,Guo-ND}:
\begin{equation}
\label{amplB0}
{\cal M}_J(s)|_{\rm N/D}=
\frac{{\cal M}_J(s)|_{\rm tree}}{1-16\pi B_0(s){\cal T}^0_J(s)|_{\rm tree}}\ ,
\end{equation}
The two-propagator Feynman integral $B_0(s)$ is nothing else but the $g(s)$ function used in the $N/D$ method~\cite{Oset-ND,Palomar-ND,Guo-ND}.
Actually, the integral is ultraviolet divergent and has a local indetermination $C$
($a^{SL}(s_0)$ in the $N/D$ analysis~\cite{Guo-ND}).
In a quantum field theory framework, this divergence is made finite by means of the corresponding coupling in the mesonic Lagrangian.

\section{\mbox{\boldmath $\eta$}-\mbox{\boldmath $\eta^\prime$} mixing}
\label{app.eta-mix}
\subsection{State mixing}
Let us consider the two-dimensional space of isoscalar pseudoscalar mesons.
We collect the $SU(3)$ octet and singlet fields in the doublet $\eta_B^T\equiv (\eta_8,\eta_1)$.
The quadratic term in the Lagrangian takes the form
\begin{equation}
\label{mixingLag}
{\cal L}=\frac{1}{2}\partial_\mu\eta_B^T{\cal K}\partial^\mu\eta_B-
\frac{1}{2}\eta_B^T{\cal M}^2\eta_B\ ,
\end{equation}
with
\begin{equation}
{\cal K}=\left(
\begin{array}{cc}
1+\delta_8 & \delta_{81}\\
\delta_{81} & 1+\delta_1\\
\end{array}
\right)\ ,\qquad
{\cal M}^2=\left(
\begin{array}{cc}
M_8^2 & M_{81}^2\\
M_{81}^2 & M_1^2\\
\end{array}
\right)\ .
\end{equation}
The mass matrix elements are defined as
\begin{equation}
\begin{array}{rcl}
M_8^2&=&\stackrel{\circ}{M_8^2}+\Delta M_8^2\ ,\\
M_1^2&=&M_0^2+\stackrel{\circ}{M_1^2}+\Delta M_1^2\ ,\\
M_{81}^2&=&\stackrel{\circ}{M_{81}^2}+\Delta M_{81}^2\ ,
\end{array}
\end{equation}
where $M_0^2$ denotes the $U(1)_A$ anomaly contribution to the $\eta_1$ mass,
$\stackrel{\circ}{M_i^2} (i=8,1)$ the ${\cal O}(\delta^0)$ quark-mass contributions to the octet and singlet isoscalar masses,
\begin{equation}
\stackrel{\circ}{M_8^2}=\frac{1}{3}(4\stackrel{\circ}{M_K^2}-\stackrel{\circ}{M_\pi^2})\ ,\qquad
\stackrel{\circ}{M_1^2}=\frac{1}{3}(2\stackrel{\circ}{M_K^2}+\stackrel{\circ}{M_\pi^2})\ ,
\end{equation}
and
$\stackrel{\circ}{M_{81}^2}=-\frac{2\sqrt{2}}{3}(\stackrel{\circ}{M_K^2}-\stackrel{\circ}{M_\pi^2})$,
with $\stackrel{\circ}{M_K^2}$ and $\stackrel{\circ}{M_\pi^2}$ the kaon and pion masses at
${\cal O}(\delta^0)$ in the combined chiral and $1/N_C$ expansion.
$\Delta M_8^2$, $\Delta M_1^2$ and $\Delta M_{81}^2$ modify the lowest order values of the
mass-matrix elements.

To first order in $\delta_8$, $\delta_1$ and $\delta_{81}$, the kinetic matrix ${\cal K}$ can be diagonalised through the following field redefinition\footnote{
In general $Z^{1/2}\cdot {\cal K}\cdot {Z^{1/2}}^\dag=I_2$ but in the case of large-$N_C$ ChPT
the matrix $Z^{1/2}$ is real since chiral loops start at ${\cal O}(\delta^2)$ and are not considered.}:
\begin{equation}
\begin{array}{c}
\eta_B={Z^{1/2}}^T\cdot\hat\eta\equiv {Z^{1/2}}^T\cdot\left(
\begin{array}{c}
\hat\eta_8\\
\hat\eta_1
\end{array}
\right)\ ,\qquad
Z^{1/2}\cdot {\cal K}\cdot {Z^{1/2}}^T=I_2\ ,\\[4ex]
Z^{1/2}=\left(
\begin{array}{cc}
1-\delta_8/2    & -\delta_{81}/2\\
-\delta_{81}/2 & 1-\delta_1/2
\end{array}
\right)\ .
\end{array}
\end{equation}
In the $\hat\eta$ basis the mass matrix takes the form
\begin{equation}
\label{etaBtoetahat}
\widehat {\cal M}^2=Z^{1/2}\cdot {\cal M}^2\cdot {Z^{1/2}}^T\ ,\
\end{equation}
where
\begin{equation}
\begin{array}{rcl}
\widehat M_8^2&=&\stackrel{\circ}{M_8^2}(1-\delta_8)-
                                    \stackrel{\circ}{M_{81}^2}\delta_{81}+\Delta M_8^2\ ,\\
\widehat M_1^2&=&(M_0^2+\stackrel{\circ}{M_1^2})(1-\delta_1)-
                                    \stackrel{\circ}{M_{81}^2}\delta_{81}+\Delta M_1^2\ ,\\
\widehat M_{81}^2&=&\stackrel{\circ}{M_{81}^2}(1-(\delta_8+\delta_1)/2)-
                                         (M_0^2+\stackrel{\circ}{M_8^2}+\stackrel{\circ}{M_1^2})\delta_{81}/2+
                                          \Delta M_{81}^2\ ,\\[1ex]
\end{array}
\end{equation}
to first order in $\Delta M^2$ (and products $\delta\times\Delta M^2$).

The physical mass eigenstates are obtained after diagonalising the matrix $\widehat {\cal M}^2$ with an orthogonal transformation
\begin{equation}
\label{etahattoetaP}
\begin{array}{c}
\widehat {\cal M}^2=R^T\cdot {\cal M}_D^2\cdot R\ ,\qquad
\hat\eta=R^T\cdot\eta_P\equiv R^T\cdot\left(
\begin{array}{c}
\eta\\
\eta^\prime
\end{array}
\right)\ ,\\[2ex]
R\equiv\left(
\begin{array}{cc}
\cos\theta_P & -\sin\theta_P\\
\sin\theta_P  & \cos\theta_P
\end{array}
\right)\ .
\end{array}
\end{equation}
One gets the relations
\begin{equation}
M_\eta^2+M_{\eta^\prime}^2=\widehat M_8^2+\widehat M_1^2\ ,\qquad
M_{\eta^\prime}^2-M_\eta^2=\sqrt{(\widehat M_8^2-\widehat M_1^2)^2+4\widehat M_{81}^2}\ ,
\end{equation}
and
\begin{equation}
\tan\theta_P=
\frac{\widehat M_8^2-M_\eta^2}{\widehat M_{81}^2}=
\frac{M_{\eta^\prime}^2-\widehat M_1^2}{\widehat M_{81}^2}=
\frac{\widehat M_{81}^2}{M_{\eta^\prime}^2-\widehat M_8^2}=
\frac{\widehat M_{81}^2}{\widehat M_1^2-M_\eta^2}\ .
\end{equation}

Therefore, the matrix from the bare to the physical basis is given by
$\eta_B=(R\cdot Z^{1/2})^T\cdot\eta_P$
with
\begin{equation}
\label{NLOmixing}
R\cdot Z^{1/2}=\left(
\begin{array}{lr}
\cos\theta_P(1-\delta_8/2)+\sin\theta_P\delta_{81}/2 &\
-\sin\theta_P(1-\delta_1/2)-\cos\theta_P\delta_{81}/2 \\[1ex]
\sin\theta_P(1-\delta_8/2)-\cos\theta_P\delta_{81}/2 &\
\cos\theta_P(1-\delta_1/2)-\sin\theta_P\delta_{81}/2
\end{array}
\right)\ .
\end{equation}

In large-$N_C$ ChPT,
\begin{equation}
\label{deltaLN}
\delta_8=\frac{8 L_5}{f^2}\stackrel{\circ}{M_8^2}\ ,\qquad
\delta_1=\frac{8 L_5}{f^2}\stackrel{\circ}{M_1^2}+\Lambda_1\ ,\qquad
\delta_{81}=\frac{8 L_5}{f^2}\stackrel{\circ}{M_{81}^2}\ ,
\end{equation}
and
\begin{equation}
\begin{array}{rcl}
\Delta M_8^2 &=&
\displaystyle
{16 L_8 \over f^2}\left(\stackrel{\circ}{M_8^4}+\stackrel{\circ}{M_{81}^4}\right)\ ,\\[2ex]
\Delta M_1^2 &=&
\displaystyle
{16 L_8 \over f^2}\left(\stackrel{\circ}{M_1^4}+\stackrel{\circ}{M_{81}^4}\right)\ ,\\[2ex]
\Delta M_{81}^2 &=&
\displaystyle
2\!\stackrel{\circ}{M_{81}^2}\left({16 L_8\over f^2}\!\stackrel{\circ}{M_K^2}+\Lambda_2\right)\ .
\end{array}
\end{equation}

In RChT,
\begin{eqnarray}
\delta_8=\frac{8 c_d c_m}{M_S^2}\frac{\stackrel{\circ}{M_8^2}}{f^2}\ ,\qquad
\delta_1=\frac{8 c_d c_m}{M_S^2}\frac{\stackrel{\circ}{M_1^2}}{f^2}\ ,\qquad
\delta_{81} =\frac{8 c_d c_m}{M_S^2}\frac{\stackrel{\circ}{M_{81}^2}}{f^2}\ ,
\end{eqnarray}
and
\begin{equation}
\begin{array}{rcl}
\Delta M_8^2 &=&
\displaystyle
\frac{8  c_m^2}{M_S^2 f^2}
\left(\stackrel{\circ}{M_8^4}+\stackrel{\circ}{M_{81}^4}\right)\ ,\\[2ex]
\Delta M_1^2 &=&
\displaystyle
\frac{8  c_m^2}{M_S^2 f^2}\left(\stackrel{\circ}{M_1^4}+\stackrel{\circ}{M_{81}^4}\right)\ ,\\[2ex]
\Delta M_{81}^2 &=&
\displaystyle
\frac{16  c_m^2}{M_S^2 f^2}\stackrel{\circ}{M_{81}^2}\stackrel{\circ}{M_K^2}\ .
\end{array}
\end{equation}
The low-energy constants of large-$N_C$ ChPT are related to the RChT couplings
(if higher-mass pseudoscalar resonance contributions are neglected) through
$L_5=c_d c_m/M_S^2$ and $L_8=c_m^2/2 M_S^2$ \cite{RChTa}.

\subsection{Decay constants}
The decay constants in the $\eta$-$\eta^\prime$ system are defined as matrix elements of axial currents
(at the quark level $A_\mu^a\equiv\bar q\gamma_\mu\gamma_5\frac{\lambda^a}{\sqrt{2}}q$)
\begin{equation}
\label{decayconstants}
\langle 0|A_\mu^a(0)|P(p)\rangle=i\sqrt{2}f_P^a p_\mu\quad (a=8,0; P=\eta,\eta^\prime)\ .
\end{equation}
Each of the two mesons has both, octet and singlet components.
Consequently, Eq.~(\ref{decayconstants}) defines \emph{four independent} decay constants, $f_P^a$.
In the convention of Ref.~\cite{Leutwyler:1997yr},
\begin{equation}
\{f_P^a\}=\left(
\begin{array}{cc}
f^8_\eta                 & f^0_\eta \\[1ex]
f^8_{\eta^\prime} & f^0_{\eta^\prime}
\end{array}
\right)=\left(
\begin{array}{lr}
f_8\cos\theta_8 & -f_0\sin\theta_0 \\[1ex]
f_8\sin\theta_8  &  f_0\cos\theta_0
\end{array}
\right)\ .
\end{equation}

In the chiral Lagrangian formalism, the physical masses and decay constants are obtained from the part of the effective action quadratic in the nonet fields with the correlator of two axial currents,
the former from the location of the poles, the latter from their residues.
Following the procedure outlined in Ref.~\cite{GL85},
the decay constants are given by $f_P^a=f[(F^\dag)^{-1}{\cal K}]_P^a$,
where $f=f_\pi=92.2$ MeV at ${\cal O}(\delta^0)$ and $F$ is the matrix which simultaneously diagonalises the kinetic and mass matrices of the Lagrangian in Eq.~(\ref{mixingLag})
\begin{equation}
{\cal K}=F^\dag I_2 F\ ,\qquad {\cal M}^2=F^\dag {\cal M}_D^2 F\ .
\end{equation}
From Eqs.~(\ref{etaBtoetahat},\ref{etahattoetaP}), we get $F=R\cdot ({Z^{1/2}}^T)^{-1}$ and then
$f_P^a=f [R\cdot ({Z^{1/2}}^T)^{-1}]_P^a$.
The matrix $F$ is exactly the same that relates the physical and the bare fields,
$\eta_P=R\cdot ({Z^{1/2}}^T)^{-1}\cdot\eta_B$.

To first order in $\delta_8$, $\delta_1$ and $\delta_{81}$, the $f_P^a$ are written as
\begin{equation}
\begin{array}{l}
f^8_\eta/f=\cos\theta_P(1+\delta_8/2)-\sin\theta_P\delta_{81}/2\ , \\[1ex]
f^0_\eta/f=-\sin\theta_P(1+\delta_1/2)+\cos\theta_P\delta_{81}/2\ , \\[1ex]
f^8_{\eta^\prime}/f=\sin\theta_P(1+\delta_8/2)+\cos\theta_P\delta_{81}/2 \ , \\[1ex]
f^0_{\eta^\prime}/f=\cos\theta_P(1+\delta_1/2)+\sin\theta_P\delta_{81}/2 \ ,
\end{array}
\end{equation}
and the two basic decay constants and two angles as
$f_8=f(1+\delta_8/2)$, $f_0=f(1+\delta_1/2)$,
$\theta_8=\theta_P+\arctan(\delta_{81}/2)$, and $\theta_0=\theta_P-\arctan(\delta_{81}/2)$,
respectively.



\begin{thebibliography}{99}

\bibitem{GL85}
  J.~Gasser and H.~Leutwyler,
  Nucl.\ Phys.\  B {\bf 250} (1985) 465.

\bibitem{Pich:1995bw}
  A.~Pich,
  Rept.\ Prog.\ Phys.\  {\bf 58} (1995) 563
  [arXiv:hep-ph/9502366].
  G.~Ecker,
  Prog.\ Part.\ Nucl.\ Phys.\  {\bf 35} (1995) 1
  [arXiv:hep-ph/9501357].

\bibitem{U3-ChPT}
  R.~Kaiser and H.~Leutwyler,
  Eur.\ Phys.\ J.\ C {\bf 17}, 623 (2000)
 [hep-ph/0007101].

\bibitem{RChTb}
  G.~Ecker, J.~Gasser, H.~Leutwyler, A.~Pich and E.~de Rafael,
  Phys.\ Lett.\  B {\bf 223} (1989) 425.

\bibitem{GAMSexp}
  A.~M.~Blik {\it et al.},
  Phys.\ Atom.\ Nucl.\  {\bf 72} (2009) 231
  [Yad.\ Fiz.\  {\bf 72} (2009) 258].

\bibitem{VESexp}
  V.~Dorofeev {\it et al.},
  Phys.\ Lett.\  B {\bf 651} (2007) 22
  [arXiv:hep-ph/0607044].

\bibitem{Alde:1986nw}
  D.~Alde {\it et al.}
  Phys.\ Lett.\  B {\bf 177}, 115 (1986).

\bibitem{Briere:1999bp}
  R.~A.~Briere {\it et al.}  [CLEO Collaboration],
  Phys.\ Rev.\ Lett.\  {\bf 84}, 26 (2000)
  [arXiv:hep-ex/9907046].

\bibitem{Venanzoni:2010ix}
  G.~Venanzoni [ for the KLOE-2 Collaboration ],
  Chin.\ Phys.\  {\bf C34 } (2010)  918-923.
  [arXiv:1001.3591 [hep-ex]].

\bibitem{Li:2009jd}
  H.~B.~Li,
  J.\ Phys.\ G {\bf 36}, 085009 (2009)
  [arXiv:0902.3032 [hep-ex]].

\bibitem{Fariborz:1999gr}
  A.~H.~Fariborz and J.~Schechter,
  Phys.\ Rev.\  D {\bf 60}, 034002 (1999)
  [arXiv:hep-ph/9902238].

\bibitem{Borasoy:2005du}
  B.~Borasoy and R.~Nissler,
  Eur.\ Phys.\ J.\  A {\bf 26}, 383 (2005)
  [arXiv:hep-ph/0510384].

\bibitem{Cronin:1967jq}
  J.~A.~Cronin,
  Phys.\ Rev.\  {\bf 161} (1967) 1483.

\bibitem{Schwinger:1968zz}
  J.~Schwinger,
  Phys.\ Rev.\  {\bf 167} (1968) 1432.

\bibitem{Di Vecchia:1980sq}
  P.~Di Vecchia, F.~Nicodemi, R.~Pettorino and G.~Veneziano,
  Nucl.\ Phys.\  B {\bf 181} (1981) 318.

\bibitem{Fajfer:1987ij}
  S.~Fajfer and J.~M.~Gerard,
  Z.\ Phys.\  C {\bf 42} (1989) 431.

\bibitem{HerreraSiklody:1999ss}
  P.~Herrera-Siklody,
  arXiv:hep-ph/9902446.



\bibitem{Schechter:1993tc}
  J.~Schechter and Y.~Ueda,
  Phys.\ Rev.\  D {\bf 3} (1971) 2874
  [Erratum-ibid.\  D {\bf 8} (1973) 987].

\bibitem{Singh:1975aq}
  C.~A.~Singh and J.~Pasupathy,
  Phys.\ Rev.\ Lett.\  {\bf 35} (1975) 1193
  [Erratum-ibid.\  {\bf 35} (1975) 1748].

\bibitem{Masjuanproceeding}
  P.~Masjuan,
  PoS C {\bf D09} (2009) 117
  [arXiv:0910.0140 [hep-ph]].

\bibitem{Escribanoproceeding}
  R.~Escribano,
  AIP Conf.\ Proc.\  {\bf 1257} (2010) 686
  [arXiv:1003.5228 [hep-ph]].


\bibitem{Montanet:1994xu}
  L.~Montanet {\it et al.}  [Particle Data Group],
  Phys.\ Rev.\  D {\bf 50}, 1173 (1994).

\bibitem{Nakamura:2010zzi}
  K.~Nakamura {\it et al.}  [Particle Data Group],
  J.\ Phys.\ G {\bf 37} (2010) 075021.

\bibitem{Amelin:2005rz}
  D.~V.~Amelin {\it et al.},
  Phys.\ Atom.\ Nucl.\  {\bf 68}, 372 (2005)
  [Yad.\ Fiz.\  {\bf 68}, 401 (2005)].

\bibitem{Kalbfleisch:1974ku}
  G.~R.~Kalbfleisch,
  Phys.\ Rev.\  D {\bf 10}, 916 (1974).

\bibitem{:2009tia}
  T.~K.~Pedlar {\it et al.}  [CLEO Collaboration],
  Phys.\ Rev.\  D {\bf 79} (2009) 111101
  [arXiv:0904.1394 [hep-ex]].

\bibitem{Kubis1}
    B. Kubis,
    EPJ Web Conf. {\bf 3} (2010) 01008
    [arXiv:0912.3440 [hep-ph]];
 B.~Kubis and S.~P.~Schneider,
 Eur.\ Phys.\ J.\  C {\bf 62} (2009) 511
 [arXiv:0904.1320 [hep-ph]].

\bibitem{MJJV-unita}
  P.~Masjuan, J.~J.~Sanz-Cillero and J.~Virto,
  Phys.\ Lett.\  B {\bf 668} (2008) 14
  [arXiv:0805.3291 [hep-ph]];
  J.~J.~Sanz-Cillero,
  arXiv:1002.3512 [hep-ph].


\bibitem{Oset-ND}
    J.A. Oller and E. Oset,
    Phys. Rev. D {\bf 60} (1999) 074023
    [arXiv:hep-ph/9809337];
%
  Z.~H.~Guo, J.~Prades and J.~A.~Oller,
  Nucl.\ Phys.\ Proc.\ Suppl.\  {\bf 207-208} (2010) 184.

\bibitem{Palomar-ND}
    J.A. Oller,  E. Oset and J.E. Palomar,
    Phys. Rev. D {\bf 63} (2001) 114009
    [arXiv:hep-ph/0011096].

\bibitem{Guo-ND}
  Z.~H.~Guo and J.~A.~Oller,
  arXiv:1104.2849 [hep-ph].

\bibitem{LeutwylerBounds}
  H.~Leutwyler,
  Phys.\ Lett.\  B {\bf 374} (1996) 163
  [arXiv:hep-ph/9601234].


\bibitem{LKdecays}
  R.~Kaiser and H.~Leutwyler,
  arXiv:hep-ph/9806336.

\bibitem{BijnensEtas}
  J.~Bijnens,
  Acta Phys.\ Slov.\  {\bf 56} (2006) 305
  [arXiv:hep-ph/0511076].

\bibitem{Ambrosino:2009sc}
  F.~Ambrosino, A.~Antonelli, M.~Antonelli, F.~Archilli, P.~Beltrame, G.~Bencivenni, S.~Bertolucci, C.~Bini {\it et al.},
  JHEP {\bf 0907 } (2009)  105.
  [arXiv:0906.3819 [hep-ph]].

\bibitem{Bijnens-anomalous}
  J.~Bijnens,
  Int.\ J.\ Mod.\ Phys.\  A {\bf 8} (1993) 3045.

\bibitem{Siklodyetaetapmixing}
  P.~Herrera-Siklody, J.~I.~Latorre, P.~Pascual and J.~Taron,
  Nucl.\ Phys.\  B {\bf 497} (1997) 345
  [arXiv:hep-ph/9610549];
%
  Phys.\ Lett.\  B {\bf 419} (1998) 326
  [arXiv:hep-ph/9710268].

\bibitem{EFTPich}
  A.~Pich,
  arXiv:hep-ph/9806303.


\bibitem{Bijnensg2}
  J.~Bijnens, E.~Pallante and J.~Prades,
  Phys.\ Rev.\ Lett.\  {\bf 75} (1995) 1447
  [Erratum-ibid.\  {\bf 75} (1995) 3781]
  [arXiv:hep-ph/9505251].

\bibitem{Gasser-ggtopipi}
  J.~Gasser, M.~A.~Ivanov and M.~E.~Sainio,
  Nucl.\ Phys.\  B {\bf 728} (2005) 31
  [arXiv:hep-ph/0506265];
%
  J.~Gasser, M.~A.~Ivanov and M.~E.~Sainio,
  Nucl.\ Phys.\  B {\bf 745} (2006) 84
  [arXiv:hep-ph/0602234].

\bibitem{Leutwyler:1997yr}
  H.~Leutwyler,
  Nucl.\ Phys.\ Proc.\ Suppl.\  {\bf 64} (1998) 223
  [arXiv:hep-ph/9709408].


\bibitem{ChPTEcker}
  G.~Ecker,
  Prog.\ Part.\ Nucl.\ Phys.\  {\bf 35} (1995) 1
  [arXiv:hep-ph/9501357].


\bibitem{BEGLis}
  J.~Bijnens, G.~Ecker and J.~Gasser,
  arXiv:hep-ph/9411232.

\bibitem{PichLis}
  A.~Pich,
  PoS C {\bf ONFINEMENT8} (2008) 026
  [arXiv:0812.2631 [hep-ph]].

\bibitem{PAs}
  J.~Bijnens, E.~Gamiz, E.~Lipartia and J.~Prades,
  JHEP {\bf 0304} (2003) 055
  [arXiv:hep-ph/0304222];
  S.~Peris,
  Phys.\ Rev.\  D {\bf 74} (2006) 054013
  [arXiv:hep-ph/0603190];
  P.~Masjuan and S.~Peris,
  JHEP {\bf 0705} (2007) 040
  [arXiv:0704.1247 [hep-ph]];
  P.~Masjuan and S.~Peris,
  Phys.\ Lett.\  B {\bf 663} (2008) 61
  [arXiv:0801.3558 [hep-ph]];
  P.~Masjuan Queralt,
  arXiv:1005.5683 [hep-ph].

\bibitem{RChTa}
  G.~Ecker, J.~Gasser, A.~Pich and E.~de Rafael,
  Nucl.\ Phys.\  B {\bf 321} (1989) 311.
\bibitem{RChTc}
  V.~Cirigliano, G.~Ecker, M.~Eidem\"uller, R.~Kaiser, A.~Pich and J.~Portol\'es,
  Nucl.\ Phys.\  B {\bf 753} (2006) 139
  [arXiv:hep-ph/0603205].

\bibitem{largeNC}
  G.~'t Hooft,
  Nucl.\ Phys.\  B {\bf 72} (1974) 461;
%
  Nucl.\ Phys.\  B {\bf 75} (1974) 461;
%
  E.~Witten,
  Nucl.\ Phys.\  B {\bf 160} (1979) 57.

\bibitem{cd+EoM}
    L.Y. Xiao and J.J. Sanz-Cillero,
    Phys. Lett. {\bf B 659} (2008) 452-456
    [arXiv:0705.3899 [hep-ph]].

\bibitem{tad-Bernard}
    V. Bernard, N. Kaiser and Ulf G. Meissner,
    Nucl. Phys. {\bf B 364} (1991) 283.

\bibitem{tad-Fpi}
    J.J. Sanz-Cillero,
    Phys. Rev. {\bf D 70} (2004) 094033.

\bibitem{Guo-aIJ}
    Z.H. Guo and J.J. Sanz-Cillero,
    Phys. Rev. D {\bf 79} (2009) 096006
    [arXiv:0903.0782 [hep-ph]].

\bibitem{ms-split}
    V. Cirigliano, G. Ecker, H. Neufeld and A. Pich,
    JHEP {\bf 0306} (2003) 012.


\bibitem{cdcm}
    M. Jamin, J.A. Oller and A. Pich,
    Nucl.~Phys.~{\bf B 622} (2002) 279.

\bibitem{cdcm2}
    M.~Jamin, J.A.~Oller and A.~Pich,
    Nucl.\ Phys.\  {\bf B 587} (2000) 331-362.

\bibitem{Ecker-tensor}
    G. Ecker and C. Zauner,
    Eur. Phys. J. C {\bf 52} (2007) 315-323
    [arXiv:0705.0624 [hep-ph]].


\bibitem{Escribano:2006mb}
  R.~Escribano,
  Phys.\ Rev.\  D {\bf 74} (2006) 114020
  [arXiv:hep-ph/0606314].

\bibitem{CCL-sigma}
  I.~Caprini, G.~Colangelo and H.~Leutwyler,
  Phys.\ Rev.\ Lett.\  {\bf 96} (2006) 132001
  [arXiv:hep-ph/0512364].

\bibitem{preliminary-BESIII}
  C.~Shen,
  [arXiv:1012.1377 [hep-ex]].



\bibitem{RChT-width}
    D. G\'omez-Dumm,  A. Pich and J. Portol\'es,
    Phys. Rev. D {\bf 62} (2000) 054014
    [arXiv:hep-ph/0003320];
%
    J.J. Sanz-Cillero and  A. Pich,
    Eur. Phys. J. C {\bf 27} (2003) 587-599
    [arXiv:hep-ph/0208199]

\bibitem{RGE}
    J.J. Sanz-Cillero
    Phys. Lett. B {bf 681} (2009) 100-104
    [arXiv:0905.3676 [hep-ph]].




\end{thebibliography}
\end{document}